\documentclass[useAMS,usenatbib]{mn2e}
\usepackage[dvips]{graphicx}

\usepackage{fixltx2e}
\usepackage{epstopdf}
\usepackage{times}
\usepackage[fleqn]{amsmath}
\newcommand{\aap}{A\&A}
\newcommand{\araa}{ARA\&A}
\newcommand{\mnras}{MNRAS}
\newcommand{\pasp}{PASP}
\newcommand{\apj}{ApJ}
\newcommand{\apjl}{ApJ}
\newcommand{\apjs}{ApJS}

\newcommand{\ssr}{Space Sci. Rev.}
\newcommand{\aj}{AJ}
\newcommand{\physrep}{Phys. Rep.}
\newcommand{\prd}{Phys. Rev. D}

\newcommand{\nbody}{{\sc nbody}}

\title[Black hole binaries and gravitational waves]
      {Black hole binaries in galactic nuclei and gravitational wave sources}
\author[J. Hong, \& H. M. Lee]
{Jongsuk Hong$^{1,2}$\thanks{E-mail: jshong@astro.snu.ac.kr (JH); hmlee@snu.ac.kr (HML)}
 , Hyung Mok Lee$^{1,3}$\footnotemark[1] \\
$^1$Astronomy Program, Department of Physics and Astronomy, Seoul National University , 1 Gwanak-ro, Gwanak-gu, Seoul 151-742, Korea\\
$^2$Department of Astronomy, Indiana University, Bloomington, IN 47404-7105, USA\\
$^3$Center for Theoretical Physics,  Seoul National University , 1 Gwanak-ro, Gwanak-gu, Seoul 151-742, Korea}
\begin{document}

\date{Accepted 2015 January 6.  Received 2014 December 9; in original form 2014 July 31}
\pagerange{\pageref{firstpage}--\pageref{lastpage}} \pubyear{2014}
\maketitle
\label{firstpage}

\begin{abstract}
Stellar black hole (BH) binaries are one of the most promising gravitational wave (GW) sources for GW detection by the ground-based detectors. Nuclear star clusters (NCs) located at the centre of galaxies are known to harbour massive black holes (MBHs) and to be bounded by a gravitational potential by other galactic components  such as the galactic bulge. Such an environment of NCs provides a favourable conditions for the BH-BH binary formation by the gravitational radiation capture due to the high BH number density and velocity dispersion. 
%Because the merging times of these binaries are very short compared to the cluster life time, the binary formation rates can be directly converted to the merger rates. 
We carried out detailed numerical study of the formation of BH binaries in the NCs using a series of N-body simulations for equal-mass cases. There is no mass segregation introduced. 
We have derived scaling relations of the binary formation rate with the velocity dispersion of the stellar system beyond the radius of influence and made estimates of the rate of formation of black hole binaries per unit comoving volume and thus expected detection rate by integrating the binary formation rate over galaxy population within the detection distance of the advanced detectors.
We find that the overall formation rates for BH-BH binaries per NC is $\sim10^{-10}{\rm yr}^{-1}$ for the Milky-Way-like galaxies and weakly dependent on the mass of MBH as $\Gamma\propto$ {\small $M_{\rm MBH}^{3/28}$}.  We estimate the detection rate of 0.02-14 yr$^{-1}$ for advanced LIGO/Virgo considering several factors such as the dynamical evolution of NCs, the variance of the number density of stars and the mass range of MBH giving uncertainties.  
%By implementing the post-Newtonian approximation, we also investigated the motion and the waveform of coalescing BH-BH binaries. The waveforms could differ significantly from those of the usual circular binaries since some of the gravitationally captured binaries are expected to have large eccentricities until they enter into the detection band of advanced detectors.
%For the typical BH-BH binary in Milky-Way-like galaxies, the merging time is a few days and the merging frequency is %$\sim$100Hz. 
\end{abstract}

\section{Introduction}
The gravitational wave (GW) is the propagation of ripples of the space-time curvature with speed of light.
%GWs do not interact with ordinary matter, so the detection of GW will be a good way to explore
%the vicinity of neutron stars (NS), black holes (BH), supernovae and active galactic nuclei that are too difficult to observe with electromagnetic (EM) waves.
Although  \citet{1916SPAW.......688E} predicted the existence of gravitational radiation (GR), the GWs have never
been directly detected yet. 
From 30 years observations,
\citet{2005ASPC..328...25W} found that the binary pulsar PSR 1913+16, discovered by \citet{1974ApJ...191L..59H}, exhibited the decrease of the orbital period and
the amount of decrease exactly coincide with the prediction of general relativity.
%In order to detect GWs {\it directly}, it is necessary to measure the distortion of space-time as GWs pass through.
%The first practical instruments, so-called bar detectors, were constructed by \citet{1960PhRv..117..306W},
%in order to measure the vibration of the metal bar due to the GWs.
%However, it became clear that the bar detectors are not sensitive enough to detect astrophysical signals. More sensitive detectors %based on laser interferometry have been constructed subsequently 
%(e.g., LIGO, Virgo, GEO 600 and TAMA 300).
%The GWs can be detected by neasuring the variation of the relative arm lengths as a function of time.
%The strain amplitudes of astronomical GWs are typically very small, $h\le 10^{-22}$.
%The sensitivity of ground-based GW detectors depends on the length of baselines and 
%is limited by various noises: shot, seismic, thermal noises.
%For the initial LIGO and Virgo, the theoretical sensitivity is comparable to $h\sim10^{-22}$ 
%at frequency around $f\sim 100$ Hz.
%It is expected that the sensitivity will be enhanced by a factor of 10 
%for second generation GW detectors such as advanced LIGO and Virgo.
There are several types of astronomical objects that can produce significant amount of gravitational waves: core-collapse supernovae 
\citep[e.g.,][]{1997A&A...317..140M,2010CQGra..27s4005Y}, spinning neutron stars \citep[e.g.,][]{2011GReGr..43..409A}, 
compact binary coalescences \citep[e.g.,][]{2004ApJ...601L.179K},
supermassive black holes (SMBH) \citep[e.g., binary SMBH merger,][]{2009ApJ...695..455B} 
\citep[extreme mass ratio inspirals, EMRI,][]{2006ApJ...645L.133H,2011PhRvD..84d4024M}
and cosmological density fluctuations \citep[e.g.,][]{2007PhRvD..75l3518A}.
Among those, the compact binary coalescence (CBC) involving neutron stars (NS) or stellar mass BH 
is the primary targets for the second generation of ground based detectors such as advanced LIGO and
Virgo.
Up to now, only ten NS-NS binary pulsars have been discovered in our Galaxy including PSR 1913+16, 
and a half of them will merge within a Hubble time \citep{2010ApJ...715.230O}. Based on these binaries
and theoretical studies, the rate of NS-NS merger within the horizon of the advanced detectors 
is estimated to lie between 0.4 and 100 per year \citep{2010CQGra..27q3001A}.
These NS-NS binaries are thought to have been evolved from primordial binaries. Obviously other
types such as BH-NS and BH-BH binaries can form by the evolution of binaries of high mass stars,
but their rates are highly uncertain especially because of the lack of observed systems. Since the
detection of the GWs coming from the CBC event will provide us with the information 
regarding the masses of the binaries, we should be able to distinguish different types of
binaries through the observations of GWs in the future.

Compact binaries can also be formed dynamically in stellar systems: three-body processes 
and dissipative two-body processes by tidal effect \citep{1986ApJ...310..176L} or GW emission 
\citep[hereafter GR capture,][]{1972PhRvD...5.1021H,1987ApJ...321..199Q}.
Stellar systems such as globular clusters (GC) and nuclear star clusters (NC) at the galactic nuclei 
provide good environments for the dynamical formation of compact binaries.
In GCs, when the core is dense enough, compact binaries can be formed by three-body encounters. 
These binaries become more compact through close encounters with other stars and is eventually kicked out from GCs 
when their orbital separations become very small. 
Some of the ejected binaries merge within a Hubble time in galactic field \citep{2010MNRAS.402..371B, 2011MNRAS.416..133D, 2013MNRAS.440.2714}.
On the other hand, binary formation by three-body processes is suppressed by the existence of massive BH (MBH) in galactic nuclei \citep{2004ApJ...613.1133B}.
Instead, coalescences of primordial compact binaries can be driven by the orbit coupling with central MBH, 
as known as the Kozai effect \citep{1962AJ.....67..591K}, and could occur within Hubble time \citep{2012ApJ...757...27A}. 
Compact binaries, especially BH-BH binaries, also can be formed by GR capture in NCs 
due to the high stellar density and velocity dispersion at the vicinity of MBH \citep{2009MNRAS.395.2127O}.
The black holes are expected to be highly concentrated in the central parts through the dynamical friction against 
low mass stars, providing a favorable environment for the close encounters between black holes.
%and enlarged BH number fraction by the mass segregation \citep{2009MNRAS.395.2127O}.
Such captured compact binaries usually have large eccentricities with small pericentre distance, and thus, 
their waveforms will be different from the ones produced by the circular binaries \citep{1992Sci...256..325A}. 

Numerous authors made estimates of the detection rates of GWs from CBCs with present- and planned- GW detectors using various methods:
population synthesis models for primordial binaries \citep{2004ApJ...601L.179K,2007ApJ...662..504B}, 
Monte-Carlo simulations for GCs \citep{2011MNRAS.416..133D} and Fokker-Planck simulations for NCs \citep{2009MNRAS.395.2127O}, and estimated that a few tens of events will be detected by new generation GW detectors
every year  \citep[see, for example, ][]{2010CQGra..27q3001A}.
However, most of these studies are based on simplified models and assumptions on the evolution of the stellar systems and the binaries. There remain substantial uncertainties because of  difficulties in accurately modeling of the systems with the large number of stars. 
Direct $N$-body approach is impossible for realistic systems, and therefore, 
statistical approaches such as 
Fokker-Planck models and Monte-Carlo simulations have been used, so far. 
In this paper, we focus on the binary formation of BHs by GR capture in NCs by direct $N$-body simulations.
Although, we cannot use realistic number of stars, we try to deduce important information regarding the binary formation and evolution based on scaled-down and simplified version of $N$-body simulations. Our study should provide useful
comparison with the statistical models.

This paper is organized as follows. In \S\S 2 and 3, we introduce the numerical method and the model for star clusters 
in galactic nuclei with a central MBH. The dynamical evolution of our model is presented in \S 4.
In \S 5, we describe binary formation in NCs and estimate the merger rate per galaxy. 
In \S 6, the expected detection rate for new generation GW detectors is estimated.
To give the interpretation for binary coalescences and their waveforms, 
we implement post-Newtonian approximations on the two-body motions. 
An example waveform of a binary BH coalescence in Milky-Way-like galaxies is provided in \S 7. 
Finally, we summarize in \S 8.

%\section{Methods and models}
\section{Initial models}

NCs are very dense stellar systems located at the nuclei of galaxies, regardless of the type \citep[e.g.,][]{1997AJ....114.2366C,2002AJ....123.1389B,2006ApJS..165...57C}. Their typical mass is $10^{6-7} M_{\odot}$ \citep{2005ApJ...618..237W}, the size of NCs is comparable to that of galactic GCs \citep{2004AJ....127..105B,2006ApJS..165...57C}, thus average stellar density of NCs is much higher than GCs. It is also well known that most of galaxies host MBHs at the centre \citep[e.g.,][]{1995ARA&A..33..581K,2005SSRv..116..523F}.
The coexistence and correlation of MBHs and NCs at the central region of galaxies have been studied by \citet{2009MNRAS.397.2148G}.

Observational \citep[e.g.,][]{2009A&A...502...91S} and theoretical \citep[e.g.,][]{1976ApJ...209..214B} studies for star clusters with the central MBH
showed that the density and velocity dispersion increases steeply toward the MBH, following Kepler's
profile (i.e., $\sigma\propto r^{-1/2}$).
%
% add: simulations and limits
%
The modeling of stellar systems with a central black hole has been done by numerous authors \citep[e.g.,][]{1980ApJ...242.1232Y,1984MNRAS.207..511G,1995ApJ...440..554Q,1995ApJ...446...75S,2002ApJ...567..817H}.
In order to generate $N$-body realizations for NCs with MBH, we adopt `adiabatic growth' of the MBH
as suggested by \citet{1995ApJ...446...75S}.
The MBH is assumed to grow with time as
\begin{align}
M_{\rm MBH}(t) = 
\begin{cases}
   M_{\rm MBH} \left[ 3\left({t\over t_{\rm MBH}}\right)^2-2\left({t\over t_{\rm MBH}}\right)^3 \right]  &  t \le t_{\rm MBH}\\
   M_{\rm MBH} & t > t_{\rm MBH}
\end{cases} 
\end{align}
where $M_{\rm MBH}$ and $t_{\rm MBH}$ are the final mass of MBH and the black hole growth time scale, respectively. During the growth of the MBH, the stellar system is adjusted against the potential of the MBH.
The MBH is fully grown after $t_{\rm MBH}$, and the gravitational potential of the MBH is assumed to follow that of the Plummer model
\begin{equation}
\phi_{\rm MBH}=-\frac{GM_{\rm MBH}}{\sqrt{r^2+\varepsilon_{\rm MBH}^2}}
\end{equation}
where $\varepsilon_{\rm MBH}$ is the softening parameter for avoiding unexpected effect at the singularity.
\citet{1995ApJ...446...75S} noted that $t_{\rm MBH}$ should be larger than the half-mass dynamical time 
to ensure the adiabatic growth of the MBH.
While they used the Hernquist model as the initial density distribution, we used Plummer model
with half-mass dynamical time $t_{\rm dyn,1/2}\sim 2.46$ for standard $N$-body scaling (i.e., $G=M_{\rm cl}=-4E=1$).

According to recent observations  \citep[e.g.,][]{2009A&A...502...91S} of the centre of the Milky Way 
(i.e., the vicinity of Sgr A$^*$), the velocity dispersion
of stars at about 1 parsec scale is nearly flat. This implies that the NC is almost in the isothermal state.
However, it is not possible for the isolated stellar systems to become fully isothermal. In order to realize
nearly isothermal sphere we need a very large and massive system. Instead we model the NC as 
a isothermal sphere surrounded by external potential which is assumed to be fixed. Such an external
potential could be provided  by the galactic bulge.
\citet{2011MNRAS.414.2728Y} have investigated a self-gravitating stellar systems embedded in an external potential well.
They considered a Plummer external potential,
\begin{equation}
\phi_{\rm pl}=-\frac{GM_{\rm pl}}{\sqrt{a_{\rm pl}^2+r^2}}
\end{equation}
where $M_{\rm pl}$ and $a_{\rm pl}$ are the mass and scale length of the Plummer potential, respectively.
They revealed that if the external potential well is deep enough compared to the potential of embedded stellar system, the external potential well behaves like a heat bath. 
The velocity dispersion of the embedded stellar system, therefore, becomes isothermal and there exists a quasi-equilibrium solution of a potential-density pair for the nearly isothermal stellar system \citep[see Model 2 of Fig. 10 in][]{2011MNRAS.414.2728Y}.

%In the case of NCs, the surrounding bulge can provide such a potential well. 
The bulge is one of galactic components, extending over few kpc scales.
NC and bulge are independent components of galaxies with different surface brightness profiles \citep[e.g.,][]{2003ApJ...582L..79B}.
In the Milky way, the effective radius (i.e., half-light radius) of the bulge is about 0.1 kpc, and the mass is estimated to be roughly $10^{10} M_{\odot}$.
According to a dynamical model of Galactic bulge suggested by \citet{1992ApJ...387..181K}, the kinematics of the bulge is affected by the MBH, Sgr A$^*$, at inner parsec scale, and the velocity dispersion is nearly flat from 1 to 10 parsec and increases gradually at the large radii.
%Although bulges are thought to be oblate spheroids or triaxial, 
We assume that the the bulge is a sphere in this study for the simplicity. This may not cause a serious effects on the dynamics of the nuclear cluster we are considering since the role of the bulge in our model is to confine the nuclear cluster within a few parsec.

\section{Computational methods}
In this study, we used the GPU accelerated version of \textsc{nbody6} \citep{2012MNRAS.424..545N}.
%\textsc{nbody6} code is one of the several versions of direct $N$-body code which has been developed by S. J. Aarseth for many %years.
This code includes many efficient and accurate algorithms such as the fourth-order Hermite integrator, the individual and block time steps, 
the Ahmad-Cohen neighbor scheme, the Kustaaheimo-Stiefel (KS) and chain regularization scheme \citep{1999PASP..111.1333A}.
Recently, by using numerous stream processors of GPU devices, 
calculations of gravitational interactions among stars have been significantly accelerated through massive parallelism.
Force calculation in \textsc{nbody6} code is divided into {\it regular} force from distant particles whose gravitational potential change slowly and {\it irregular} force from neighboring particles 
that give strong gravitational fluctuations. 
In the GPU accelerated version, the regular forces are calculated by GPU machines with single precision while irregular force calculation that needs better accuracy is still done by CPUs.

As we mentioned above, the external potential is composed of two parts,
\begin{equation}
\phi_{\rm ext}=\phi_{\rm MBH}+\phi_{\rm pl}
\end{equation}
where $\phi_{\rm MBH}$ is the Keplerian potential due to the central MBH. In our implementation, these external forces are added to the sum of regular forces. 

All the calculations in \nbody6 code use dimensionless time, length and mass units.
From given unit of length $\bar{r}$ in parsec and mean stellar mass $\bar{M}$ in $M_{\odot}$, the physical unit of velocity and time can be expressed as \citep{2010gnbs.book.....A}
\begin{equation}
\textrm{velocity : } 6.557\times10^{-2}\bigg( \frac{N\bar{M}}{\bar{r}}\bigg)^{1/2} \textrm{km/s},  \nonumber \\
\end{equation}
\begin{equation}
\textrm{time : } 14.94\bigg(\frac{\bar{r}^3}{N\bar{M}}\bigg)^{1/2} \textrm{Myr}
\label{eqns5}
\end{equation}
where $N$ is total number of stars.

%Since the Plummer external potential has already been implemented in \textsc{nbody6} code, 
%we added the potential of MBH in the code in a similar manner with the Plummer potential.
Table \ref{tblg1} shows model parameters of our simulations. Although galactic nuclei contain $\sim 10^6$ stars in a cubic parsec, 
it is hard to treat such a large number of particles for \textsc{nbody6} code even with GPU machine.
We, therefore, used several different number of particles ranging 
from 10,000 to 100,000 in order to build a scaling relation.
We also made several simulations with different random seeds and took average in order to reduce the statistical noises.
The masses of MBH are chosen to be 10 and 20 \% of the entire mass of cluster stars $M_{\rm cl}$ 
(but excluding the mass of the Plummer potential).
Because these masses are relatively larger than those of \citet{1995ApJ...446...75S} and \citet{2002ApJ...567..817H},
we also used longer black hole growth time scale $t_{\rm MBH}=50$. The softening parameter of the MBH is fixed to $10^{-4}$, 
which is much smaller than the radius of influence of the MBH 
(c.f., $r_{\rm inf} \equiv M_{\rm MBH}/\sigma_{*}^2 \sim 0.1$ for $M_{\rm MBH}=0.1$).
We assumed that all stars have same masses (i.e., $m=1/N$) of stellar mass BHs with the mass of $10 M_{\odot}$. 
Thus, the total mass of the cluster in physical unit becomes $M_{\rm tot} = 10NM_{\odot}$.
It is well known that there is a correlation between the mass of MBHs and the kinematics of bulges.
Recently, \citet{2013ApJ...764..184M} updated the $M_{\rm MBH}-M_{\rm bulge}$ relation and found that the mass of MBH is roughly 0.2\% of the bulge mass for $10^7 < M_{\rm MBH}/M_{\odot} < 10^{10}$ \citep[also see ][]{2015ApJ...798...54G}.
For the external potential of a bulge, we fixed the mass and scale length of the Plummer potential to 100 and 5, respectively, which corresponds to $M_{\rm MBH}/M_{\rm bulge}=0.001$ and 0.002 for our models.
Under this potential well, the embedded stellar system is expected to become nearly isothermal in a few half-mass relaxation time \citep{2011MNRAS.414.2728Y}. We also considered a model without the external Plummer potential well (Model 0) for the purpose of comparison. 

\begin{table*}
  \begin{center}
  \caption{Initial parameters for all models.}
  \begin{tabular}{l c c c c c c c c}
  \\
    \hline
    \hline
    Model & $N_{\rm cl}$ & $N_{\rm run}$ & $M_{\rm MBH}/M_{\rm cl}$ & $M_{\rm MBH}/m_{*}$ & $t_{\rm MBH}$ 
   & $\varepsilon_{\rm MBH}$ & $M_{\rm pl}$ & $a_{\rm pl}$ \\
    & (1) & (2) & (3) & (4) & (5) & (6) & (7) & (8) \\
    \hline 
    0  & 100,000 &  1 & 0.2 & 10,000 & 50 & $10^{-4}$&  -  & - \\
    \hline
    1  &  10,000 & 10 & 0.2 &  2,000 &          &    &     &   \\
    2  &  20,000 &  5 & 0.2 &  4,000 &          &    &     &   \\
    3  &  50,000 &  2 & 0.2 & 10,000 &          &    &     &   \\
    4  & 100,000 &  1 & 0.2 & 20,000 & 50 & $10^{-4}$& 100 & 5 \\ \cline{1-5}
    5  &  20,000 &  5 & 0.1 &  2,000 &          &    &     &   \\
    6  &  50,000 &  2 & 0.1 &  5,000 &          &    &     &   \\
    7  & 100,000 &  1 & 0.1 & 10,000 &          &    &     &   \\  
    \hline
    \label{tblg1}
  \end{tabular}
  \end{center}
  \begin{flushleft} Notes. (1): Number of stars in the nuclear star cluster. (2): Number of simulations with different initial random seeds. (3): MBH mass compared to the total mass of cluster. (4) Mass ratio of MBH to the stellar mass. (5): MBH growth time scale. (6): Softening parameter of MBH potential. (7): Mass of external Plummer potential. (8): Scale length of external Plummer potential.
  \end{flushleft}
\end{table*}

\section{Dynamical evolution of star clusters}
\subsection{Cluster expansion}
In order to investigate the effect of the MBH and the surrounding bulge, 
the dynamical evolution of the star cluster is presented in this section.
Here, we focus on the Model 4 with $N=100,000$ and $M_{\rm MBH}/M_{\rm cl}=0.2$.
\begin{figure}
  \centering
  \includegraphics[width=84mm]{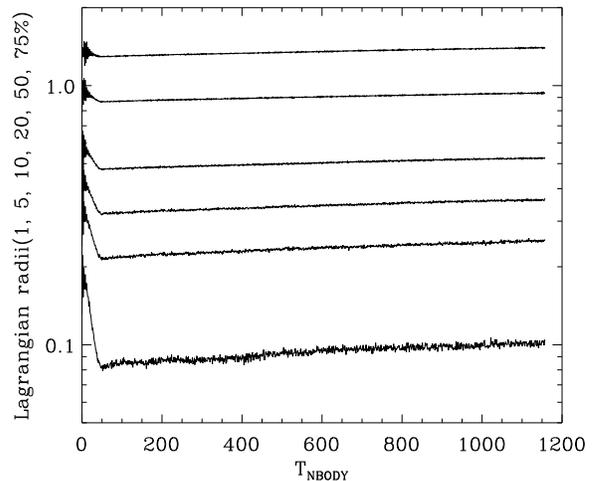}
  \caption[Time evolution of Lagrangian radii]{Time evolution of Lagrangian radii for the star cluster 
  with a growing central MBH and an external Plummer potential well. By the growth of MBH, 
  the Lagrangian radii decrease with time until $T=t_{\rm MBH}=50$. After full growth of the MBH, 
  the cluster expands due to the heating from the MBH and stars in the cusp.}
  \label{figg_lagr}
\end{figure}
\begin{figure}
  \centering
  \includegraphics[width=84mm]{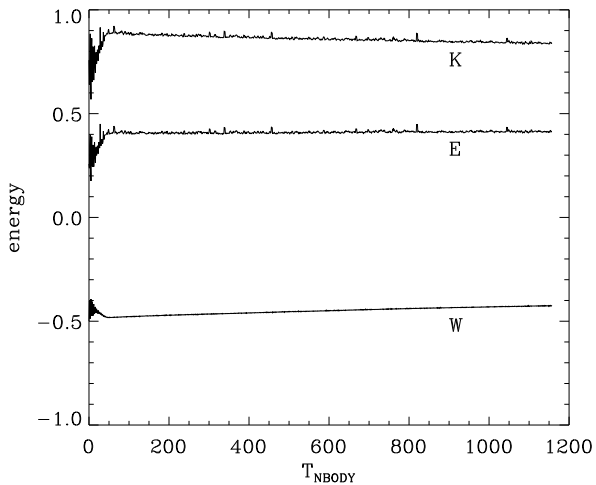}
  \caption[Time evolution of energies]{Time evolution of energies for stellar particle only. 
  To be virialized against the external potential, the kinetic energy becomes much larger than the isolated systems.
  The kinetic energy decreases with time because of the cluster expansion while the total energy is nearly conserved.}
  \label{figg_ener}
\end{figure}

Fig. \ref{figg_lagr} shows the time evolution of Lagrangian radii including 1, 5, 10, 20, 50 and 75 \% of total cluster mass. 
Before $T=t_{\rm MBH}=50$, all Lagrangian radii decrease with growth of the central MBH
to adjust with the strong potential of the MBH.
After $T=t_{\rm MBH}$, Lagrangian radii increase gradually as reported in previous studies \citep[e.g.,][]{2004ApJ...613.1133B,2006ApJ...649...91F,2012MNRAS.419...57F}.
The basic mechanism of the expansion is similar to that of post core collapse expansion of star clusters.
% However, there is no gravothermal oscillation because of the steady heating by MBH %\citep{2003gmbp.book.....H,2004ApJ...613.1133B}.
With the MBH, kinetic energy can be generated by stars in the cusp \citep{1977ApJ...217..281S}.
The MBH and the innermost star can behave like hard binaries in the core of star clusters.  Encountering other single stars in the cusp, they are to be bounded stronger and convert their internal energy to the kinetic energies of stars in the cusp.
The kinetic energies are transferred to the whole cluster via relaxation. 

While the kinetic, potential and total energies of an isolated self gravitating system are
1/4, -1/2 and -1/4 in \textsc{nbody} units, respectively, the system embedded in a potential well has larger kinetic energy than the isolated system
through the virialization \citep{2011MNRAS.414.2728Y}. 
Also, when the kinetic energy is even larger than the magnitude of the potential energy (i.e., the total energy becomes positive), 
the system will not reach the core collapse.
Fig. \ref{figg_ener} shows the time evolution of the energies. Our models are designed to have positive total energy.
Initially, the kinetic energy, potential energy of self-gravity and total energy are 0.7, -0.45 and 0.25, respectively.
However, as the black hole grows, the potential energy decreases 
because the cluster becomes more centrally concentrated as shown in Fig.  
\ref{figg_lagr} while the kinetic energy increases in response to the external potential of the bulge.
At $T=t_{\rm MBH}$, the energies become -0.5 (potential), 0.9 (kinetic) and 0.4 (total), respectively.
The potential energy increases after $T=t_{\rm MBH}$ due to the cluster expansion 
although we designed a quasi-equilibrium model by using the external Plummer potential.
Once the MBH has been grown, the numerical error is generated by the interaction between the innermost stars and the MBH at $r \to 0$. 
For Model 4, the root-mean-square (RMS) error of total energy per step $\langle \Delta E/E \rangle_{\rm rms}$ including the potential energies from external potential and MBH is $\sim10^{-3}$, overall. 
Such a large error could originate in escaping stars which are kicked out from inner region due to the less accurate force calculation 
during the regularization processes perturbed by strong gravitational potential of the MBH.
However, they have very large escaping velocity ($\sim100\sigma_*$) and rarely interact with other stars during escaping. 
When these escapers are excluded, the RMS error drops to $\sim10^{-6}$.
Although these escapers rarely interact with other stars, 
the indirect heating could cause small structural changes at $r<10^{-2}$ where escapes mostly happen, as discussed in \S 5.2.
\subsection{Radial profiles}
\begin{figure}
  \centering
  \includegraphics[width=84mm]{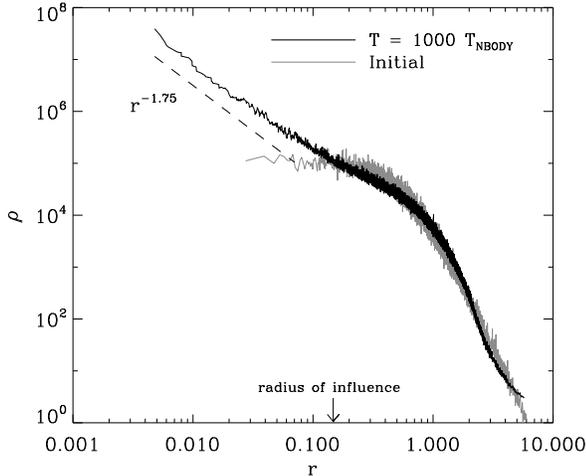}
  \caption[Density profile]{Density profiles at $T=0$ and $T=1000 T_{\rm NBODY}$. Grey line shows the initial condition.
Because of the existence of MBH, the stellar cusp, so-called Bahcall-Wolf cusp, is formed inside the radius of influence. 
Our model has less steep density profile when approaching $r_{\rm inf}$ than the theoretical profile of $r^{-1.75}$ due to the contamination from outer parts.}
  \label{figg_rho}
\end{figure}
\begin{figure}
  \centering
  \includegraphics[width=84mm]{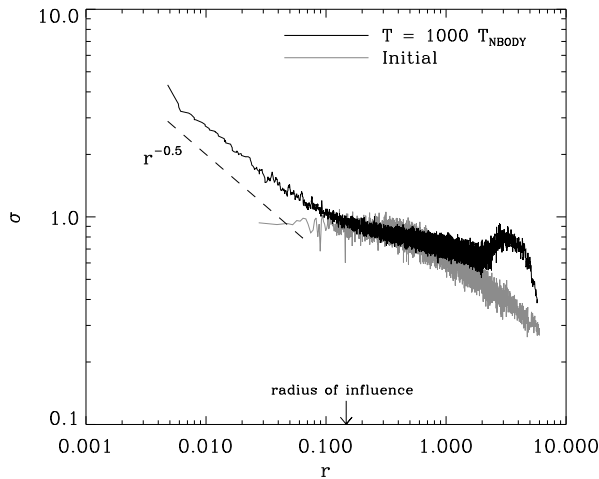}
  \caption[Velocity dispersion profile]{Velocity dispersion profiles at two epochs: $T=0$ and 1000. 
  The existence of MBH makes stars within radius of influence follow the Keplerian profile.
  For the outer parts, the velocity dispersion is flatter than the initial condition due to the external potential well.}
  \label{figg_sig}
\end{figure}

In order to describe the dynamics at the vicinity of MBH, it is very important to determine the radius of influence $r_{\rm inf}$, within which stars are directly affected by the gravitational potential of MBH. We estimated the radius of influence by following the definition of \citet{2004ApJ...613.1133B} as 
\begin{equation}
r_{\rm inf}=\frac{GM_{\rm MBH}}{2\sigma_*^2}
\end{equation}
where `2' is an empirical factor suggested by \citet{2004ApJ...613.1133B}, and $\sigma_*$ is here the density-weighted one dimensional velocity dispersion of stars 
\begin{equation}
\sigma_{*}^2\equiv\frac{\int_{r_{\rm inf}}^{r_{\rm half}} 4\pi \sigma^2(r) \rho(r) r^2 dr}
{\int_{r_{\rm inf}}^{r_{\rm half}} 4\pi \rho(r) r^2 dr}
\label{eqns7}
\end{equation}
where $r_{\rm half}$ is half-mass radius. Since $r_{\rm inf}$ and $\sigma_*$ are interrelated each other, the estimation of them is made iteratively within the relative error $10^{-4}$. 
For instance, $r_{\rm inf}=0.16$, $\sigma_*=0.79$ and the mass of stars within $r_{\rm inf}$, $M(< r_{\rm inf}) \approx 0.3$ for Model 4, respectively. 

In Fig. \ref{figg_rho}, we show the density profile of the star cluster of the Model 4 for $t=0$ and $1000 T_{\rm NBODY}$.
%The density profile is obtained at five different snapshots from $T=1000 T_{\rm NBODY}$.
%The grey line is the density profile of the initial condition.
We see in Figs. \ref{figg_lagr} and \ref{figg_ener} that the cluster is not in equilibrium but expanding.
However, as reported in \citet{2004ApJ...613.1133B}, the equilibrium profile is expected to be 
established from inner to outer regions after a few local relaxation times (c.f., $\tau_{\rm ri}\sim 100$ and $\tau_{\rm rh,0}\sim1600 T_{\rm NBODY}$ for Model 4).
The radius of influence $r_{\rm inf}$ is marked as the downward arrow. 
The slope of density cusp at $r \ll r_{\rm inf}$ is similar to -1.75 for equal-mass systems but becomes shallower at $r \to r_{\rm inf}$.
This feature has already been observed many previous studies with different approaches such as direct $N$-body \citep{2004ApJ...613.1133B} and Monte-Carlo simulations \citep{2006ApJ...649...91F}.
There is an upturn of the density near $r=5$. This radius is the scale length of the external Plummer potential, 
so stars have piled up at this radius against the external force from the potential.

The radial profile of velocity dispersion is presented in Fig. \ref{figg_sig}. The black line shows the profile at $T=1000$,
and the grey line is the initial one. 
For the outer part in Fig. \ref{figg_sig}, the velocity dispersion is not completely flat, but  flatter than the initial profile.
%When a star is ejected from the vicinity of the MBH,
%the star has very large kinetic energy and transfers its kinetic energy to other stars 
%through the interactions. Nevertheless, the velocity dispersion is flatter than the initial dispersion.
There is a bump of velocity dispersion at $r\sim5$, the scale length of the external Plummer potential.
This may have been caused by the heating effect by the external potential which is more effective where the
depth of the external potential acts like a reflecting wall. This will be more evident in the next subsection on the velocity
anisotropy. The heating through the reflection would appear mostly in the radial direction. 
When the MBH exists, stars inside $r_{\rm inf}$ are strongly affected by the MBH,
and their velocity dispersion follows the Keplerian profile (i.e., $\sigma(r) \sim r^{-0.5}$).
%One can see the slope of the velocity dispersion profile at $r<r_{\rm inf}$.
However, the slope is a little shallower than the expectation as similar to the density profile in Fig. \ref{figg_rho}. 
Another important characteristic of MBH surrounded by stars is the wandering. 
An MBH in a stellar system can move randomly, like a Brownian motion, 
by the interaction with stars which are bounded and un-bounded to the potential of the MBH.
In addition, the MBH and the innermost star can play a role as binaries in the core-collapsed star cluster.
They kick a star interacting with them with high kinetic energy, and this causes the recoil motion of the MBH.
\citet{1980ApJ...242..789L} investigated the motion of the MBH in the stellar system such as a globular cluster
and concluded that the interaction with unbound stars is the most important effect causing the motion of the MBH.
\citet{1976ApJ...209..214B} estimated the uncertainty of the position of the MBH 
by the wandering of the MBH as
\begin{equation}
r_{\rm wand}\approx 0.92 r_{\rm c} \sqrt{m_{*}/M_{\rm MBH}}
\label{eqns8}
\end{equation}
where $r_{\rm c}$ is the core radius. 
It is difficult to define the wandering radius in our simulations because there is not a well-defined core, but
\citet{2012MNRAS.419...57F} found the empirical relation for the MBH wandering radius as $r_{\rm wand} = 0.5 (m_*/M_{\rm MBH})^{0.5}$ for King models.
%Nevertheless, from the simulations with different number of stars, we confirmed the number dependence of the wandering radius.
\begin{figure}
  \centering
  \includegraphics[width=84mm]{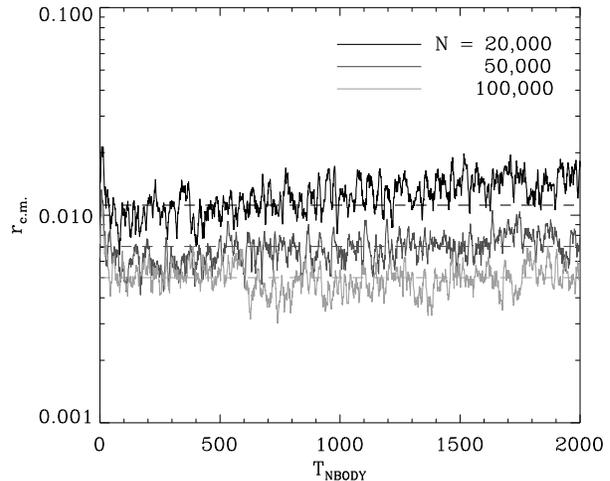}
  \caption[Wandering of MBH]{Separation between the MBH and the centre of mass of stars 
  as a function of time for simulations with different $N$. Contrasts show different numbers of stars. 
  With less number of stars, the MBH wanders more. Dashed lines represent the wondering radius
  based on the our scaling relation obtained from $N$=100,000 model.}
  \label{figg_rcm}
\end{figure}
\begin{figure}
  \centering
  \includegraphics[width=84mm]{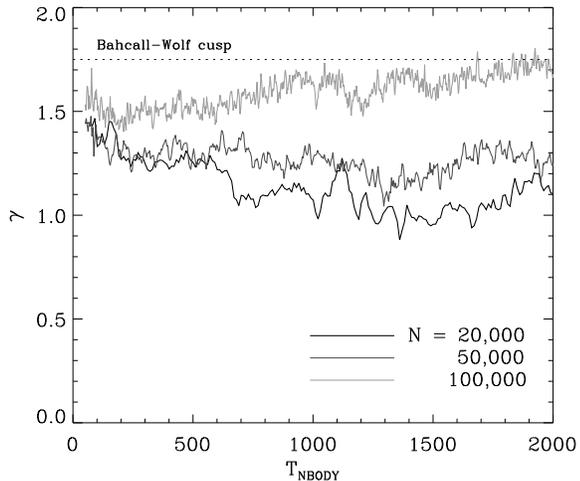}
  \caption[Time evolution of the slope of stellar cusp]{Time evolution of the slope of stellar cusp 
  within the radius of influence (see text). Different lines show results with  different numbers of stars. 
  There is a tendency of steeper slope for the larger N models. The model with $N$=100,000 reaches the Bahcall-Wolf slope at $T \sim 2000$.}
  \label{figg_gam}
\end{figure}

Fig \ref{figg_rcm} shows the wandering radius estimated by the radial distance of the MBH fixed at the origin 
to the centre of mass of the stellar particles.
The solid lines are the wandering radius in the simulations for the simulations with different number of stars ranging 
from 20,000 (top) to 100,000 (bottom) with the MBH mass of 0.2.
As the number of stars becomes larger, the wandering radius becomes smaller as shown in equation (\ref{eqns8}).
The dashed lines are the values obtained with the relation from \citet{2012MNRAS.419...57F}, but with different coefficient of 0.7 rather than 0.5. 
Thus the wondering radius follows the scaling relation with $N$ very well.
Different coefficient is likely to be from the different initial profiles.
The time evolution of the slope of the central stellar cusp $\gamma$ is shown in Fig. \ref{figg_gam}.
Due to the fact that the density cusp is smeared out by the random walk of MBH \citep{2012MNRAS.419...57F} and the density profile is contaminated by outer region at $r \to r_{\rm inf}$ as shown in Fig. \ref{figg_rho}, 
the slope is estimated with the stars at $r_{\rm wand} < r < 0.3 r_{\rm inf}$ 
(However, we used larger factor rather than 0.3 for small $N$ runs to secure enough number of stars in that range, and this may cause shallower cusp for small $N$ runs).
Different contrasts mean different number of stars. 
For larger $N$, the slope increases with time slowly toward the Bahcall-Wolf value of -1.75 
while those with smaller $N$ do not get close to the Bahcall-Wolf cusp.
For Model 4, the cusp once approaches to the Bahcall-Wolf cusp at $T=2000T_{\rm NBODY}$. This is consistent with  the result of \citet{2006ApJ...649...91F} that it takes order $10\tau_{\rm ri}$ (c.f., $\tau_{\rm ri}\sim100$) for the Bahcall-Wolf cusp to be fully-developed. However, it is still true that most of our models have shallower cusps overall.
The shallower stellar cusp will affect the merger rate of stellar mass BHs near the MBH as discussed in the next sections.

\subsection{Velocity anisotropy}
It is well known that the radial anisotropy in the velocity dispersion increases at the outer part of isolated stellar systems as a result of two-body relaxation \citep{1996MNRAS.279.1037G,1987degc.book.....S}.
The anisotropy parameter can be defined by \citep{2008gady.book.....B}
\begin{equation}
\beta \equiv 1-\frac{\sigma_{\rm t}^2}{2\sigma_{\rm r}^2}
\end{equation}
where $\sigma_{\rm t}$ and $\sigma_{\rm t}$ are the tangential and radial velocity dispersion, respectively.
This anisotropy parameter varies from  $-\infty$ (purely circular orbits) to $+1$ (purely radial orbits).
Also, in a tidal field, the radial anisotropy decreases during post core-collapse expansion due to the loss of radial orbits \citep{1997MNRAS.292..331T}.
However, in the case of stellar systems with a growing central massive object, the tangential anisotropy is developed \citep[i.e., $\beta < 0$,][]{1980ApJ...242.1232Y,1984MNRAS.207..511G,1995ApJ...440..554Q,1995ApJ...446...75S,2002ApJ...567..817H}.
\citet{1995ApJ...440..554Q} have revealed that the aspect of anisotropy is affected by the initial models. For models with a core such as isothermal sphere and isochrone model, the velocity distribution becomes isotropic in the limit $r\to0$ \citep{1980ApJ...242.1232Y,1995ApJ...440..554Q}. On the other hand, for `two-power' models for galaxies like Dehnen's models \citep{1993MNRAS.265..250D}, the velocity distribution is still tangentially biased at $r\to0$ \citep{1995ApJ...440..554Q,1995ApJ...446...75S,2002ApJ...567..817H}. 
The tangential anisotropy reaches the peak at $r \approx r_{\rm inf}$, and the velocity distribution becomes isotropic at $r \gg r_{\rm inf}$ again \citep[see Figs. 2-5 of][]{1995ApJ...440..554Q}.

\begin{figure}
  \centering
  \includegraphics[width=84mm]{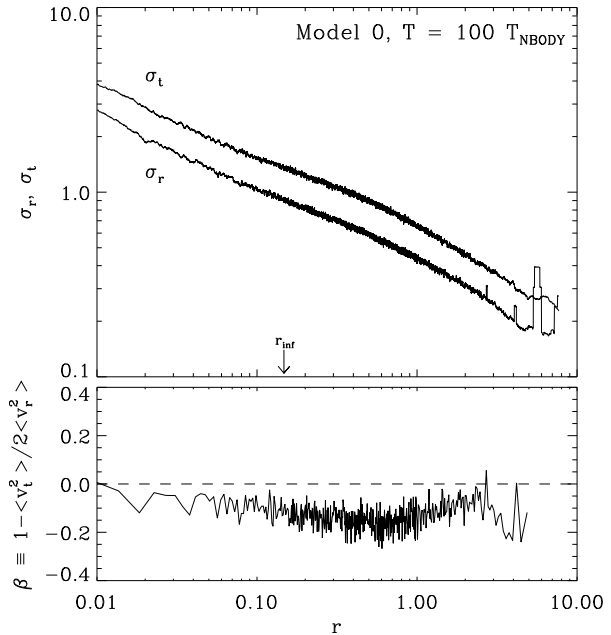}
  \caption[Velocity anisotropy without the external potential well]{Profiles of Radial and tangential velocity dispersion (upper) and the anisotropy parameter (lower) for Model 0 without the external Plummer potential. The velocity distribution is tangentially biased, and the profile is similar to that of the isochrone model in \citet{1995ApJ...440..554Q} except for the position of maximum anisotropy.}
  \label{figg_ainp}
\end{figure}
\begin{figure}
  \centering
  \includegraphics[width=84mm]{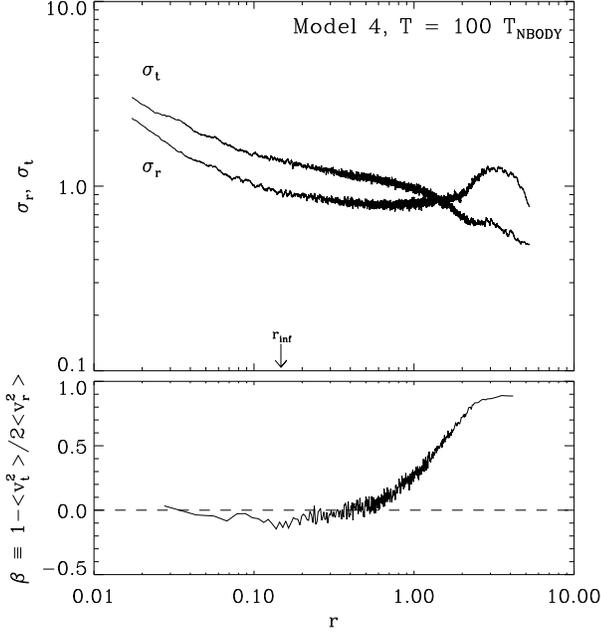}
  \caption[Velocity anisotropy with the external potential well]{Profiles of Radial and tangential velocity dispersion (upper) and the anisotropy parameter (lower) for Model 4 with the external Plummer potential. For the inner region, the velocity distribution is similar to that of isolated models. However, the tangential anisotropy is driven by the external potential well at the outer region.}
  \label{figg_aip}
\end{figure}

The radial and tangential velocity dispersion and the velocity anisotropy parameter after the growth of MBH are shown in Fig. \ref{figg_ainp} for the Model 0 without the external Plummer potential and Fig \ref{figg_aip} for the Model 4 with the potential, respectively. 
For the Model 0, the velocity dispersion decrease rapidly with radius unlike those of the model with the potential well. Although we use the Plummer model as the initial model, the anisotropy parameter is likely to be similar to that of the isochrone model as shown in \citet{1995ApJ...440..554Q}. However, the maximum anisotropy is located at the larger radius than $r_{\rm inf}$.
 On the other hand, for the Model 4, the radial velocity dispersion is enhanced due to the radial acceleration from the external potential as mentioned in the previous section. Obviously this is an artifact of the fixed external potential. If the external 
potential is a live one, the velocity dispersion and the anisotropy would be more smooth. However, as we will see in the following
section, the details of the velocity profiles and anisotropy would not affect on the estimation of binaries as the
binary formation rate at the outer part becomes very low. 
% Therefore, the anisotropy parameter becomes almost 1 at $r\approx a_{\rm pl}$. For the inner region, the velocity dispersion is %slightly tangentially biased as in the case of the isolated models.

\section{Black hole binaries}
\subsection{Close encounters and GR capture \label{subsec5.1}}
In order to estimate the merger and detection rates of BH-BH binary coalescences, 
we need to know binary formation rates as well as the orbital parameter distribution just after the capture. 
We only consider the direct capture by gravitational radiation during close encounters.
%There are several binary formation mechanisms known: primordial, dynamical three-body process, dissipative two-body %processes.
%\citet{2012ApJ...757...27A} have discussed how primordial BH-BH binaries can evolve to GW sources 
%in galactic nuclei by the secular Kozai effect and estimated the merger fraction. 
%In addition, previous studies \citep[e.g.,][]{2004ApJ...613.1133B} noted that binary formation by three-body process 
%is suppressed by the strong gravitational field at the vicinity of MBH.
%There are two of dissipative processes for two-body encounters, tidal force and GR.
%The tidal interaction is not our interest because this process is only valid for objects with finite sizes
%such as main-sequence (MS) or giant stars.

The energy loss and change of orbits by GW for binary systems were first studied by 
\citet{1963PhRv..131..435P} based on the post-Newtonian (PN) approximation. 
Later, \citet{1972PhRvD...5.1021H} extended the study of \citet{1963PhRv..131..435P} to the hyperbolic encounters.
With given masses $m_1, m_2$, a semi-major axis $a$ (defined as $a=Gm_1m_2/2E_0$ where $E_0$ is the initial orbital energy) and an eccentricity $e$, 
the energy and orbital angular momentum losses by GR are given by
\begin{align}
\Delta E&=- \frac{2}{15}\frac{G^{7/2}}{c^{5}}\frac{m_{1}^{2}m_{2}^{2}(m_{1}+m_{2})^{1/2}}{a^{7/2}(e^{2}-1)^{7/2}}\nonumber\\
&\times\biggl[(\pi-\theta_{0})(96+292e^{2}+37e^{4})+\frac{1}{3}e \sin \theta_{0}(602+457e^{2})\biggr],
\label{eqns10} 
\end{align}
\begin{equation}
\Delta L_z=- \frac{8}{5}\frac{G^{3}}{c^{5}}\frac{m_{1}^{2}m_{2}^{2}}{a^{2}(e^{2}-1)^{2}}
\biggl[(\pi-\theta_{0})(8+7e^{2})+e \sin \theta_{0}(13+e^{2})\biggr]
\label{eqns11}
\end{equation}
where $G, c$ and $\theta_{0}$ are gravitational constant, speed of light and the incidence angle at infinity
defined as $\theta_{0}=\cos^{-1}(1/e)$, respectively.
Two encountering but unbound stars, therefore, become a binary if the energy loss by GR
is larger than the orbital energy $E_0$. 
From equations (\ref{eqns10}) and (\ref{eqns11}), one can obtain the semi-major axis and the eccentricity of the captured binary as
\begin{align}
a' &=-\frac{Gm_1m_2}{2(E_0+\Delta E)} \quad  \\
e' &=\sqrt{1-\frac{(m_1+m_2)(L_{z,0}+\Delta L_z)}{Gm_1^2m_2^2a'}}
\end{align}
where the subscript 0 indicates the initial value.

%\subsection{GR capture}\label{sec_ert}
Many previous studies have discussed the GR captures of compact stars in dense stellar systems 
\citep[e.g.,][]{1987ApJ...321..199Q,1989ApJ...343..725Q,2009MNRAS.395.2127O}. 
The starting point is the cross section for GR capture.
\citet{1987ApJ...321..199Q} deduced the capture cross section under the parabolic approximation.
This approximation is valid because trajectories of the stars near the pericentre, 
where bulk of the GWs is radiated, are almost identical to parabolic with the same pericentre distance.
The equation (\ref{eqns10}) is rewritten with parabolic approximation as 
\begin{equation}
\Delta E=- \frac{85\pi}{12\sqrt{2}}\frac{G^{7/2}}{c^{5}}\frac{m_{1}^{2}m_{2}^{2}(m_{1}+m_{2})^{1/2}}{r_{\rm p}^{7/2}}
\label{eqns14}
\end{equation}
where $r_{\rm p}$ is the pericentre distance.
Again, the GR capture will happen when the energy loss is lager than the orbital energy.
By the requirement of $|\Delta E| > m_1m_2v_{\infty}^2/2(m_1+m_2)$, \citet{1989ApJ...343..725Q} obtained the maximum pericentre distance
for GR capture:
\begin{equation}
r_{\rm p,max}=\Bigg[\frac{85\pi\sqrt{2}}{12}\frac{G^{7/2}}{c^{5}}\frac{m_{1}m_{2}(m_{1}+m_{2})^{3/2}}{v_{\infty}^2}\Bigg]^{2/7}
\label{eqns15}
\end{equation}
where $v_{\infty}$ is the relative velocity at infinity. Therefore, the capture cross section is given as
\begin{align}
\Sigma_{\rm cap} &=\pi r_{\rm p,max}^2\Bigg[1+\frac{2G(m_1+m_2)}{r_{\rm p,max}v_{\infty}^2}\Bigg]
\simeq 17\frac{G^2m_1m_2\eta^{-5/7}}{c^{10/7}v_{\infty}^{18/7}}
\label{eqns16}
\end{align}
where $\eta$ is the symmetric mass ratio defined as $\eta \equiv m_1m_2/(m_1+m_2)^2$.
We assumed that gravitational focusing is dominant compared to the geometrical cross section 
for the last equality in the above equation.

%\subsection{Distribution of semi-major axis, eccentricity and merging time \label{sec_dis}}
%We presented the criteria for the formation of binaries by GR.
Here, we are going to introduce a statistical interpretation of GR captures to
understand the situations and predict the BH-BH binary coalescences in realistic regime.
We can assume that the motions of stars follow the one-dimensional normal distribution
with a given velocity dispersion $\sigma$.
From the equation (\ref{eqns16}), the distribution of the pericentre distance of encountering stars also becomes uniform
(i.e., $dS = d(\pi b^2) \simeq 2\pi G(m_1+m_2)/v_{\infty}^{2} \cdot d r_{\rm p}$) if the gravitational focusing dominates.
Therefore, for unbound close encounters that lead to the formation of binaries by GR, 
$r_{\rm p}/r_{\rm p,max}$ is distributed uniformly in the range [0, 1].
Under the parabolic approximation, the semi-major axis and the eccentricity can be rewritten with $\sigma$ and $r_{\rm p}/r_{\rm p,max}$ from the equations (\ref{eqns14}) and (\ref{eqns15}) as
\begin{equation}
a=\frac{GM}{2\sigma^2}\approx1.58\textrm{AU}\Bigg(\frac{M}{20 M_{\odot}}\Bigg) \Bigg(\frac{\sigma}{75 \textrm{km/s}}\Bigg)^{-2},
\end{equation}
\begin{align}
e &=1+\Bigg(\frac{170\pi\eta}{3}\Bigg)^{2/7} \Bigg(\frac{\sigma}{c}\Bigg)^{10/7}\Bigg(\frac{r_{\rm p}}{r_{\rm p,max}}\Bigg)\nonumber\\
&\approx 1+ 1.57 \times 10^{-5} \eta^{2/7} \Bigg(\frac{\sigma}{75 \textrm{km/s}}\Bigg)^{10/7}
\end{align}
where $M$ is the sum of masses, and we set $v_{\rm inf}=\sqrt{2} \sigma$.
By assuming $r_{\rm p}/r_{\rm p,max}$ = 1/2  typical pericentre distance for GR capture of encountering two 10$M_{\odot}$ BHs 
is 2514km, corresponding to about 840 $R_{\rm Sch}$ (Schwarzschild radius),
in Milky-Way-like galaxies (i.e., $\sigma=75$km/s).
Because the pericentre distance is nearly the same before and after capture, 
the semi-major axis and eccentricity of the binary formed by GR capture are given by
\begin{align}
a' &=a\Bigg[\Bigg(\frac{r_{\rm p}}{r_{\rm p,max}}\Bigg)^{-7/2}-1\Bigg]^{-1}\nonumber \\
&\approx0.153\textrm{AU}
\Bigg(\frac{M}{20 M_{\odot}}\Bigg) \Bigg(\frac{\sigma}{75 \textrm{km/s}}\Bigg)^{-2},
\label{eqns19}
\end{align}
\begin{align}
e'& =1-\frac{a}{a'}(e-1)\nonumber \\
&\approx 1 - 1.62 \times 10^{-4} \eta^{2/7} \Bigg(\frac{\sigma}{75 \textrm{km/s}}\Bigg)^{10/7}.
\label{eqns20}
\end{align}
For the case with velocity dispersion of $\sigma\sim75$ km/s, 
the eccentricity of a typical binary formed by GR capture is $1-e' \sim 10^{-4}$.
% which can show us the validity of the parabolic approximation again.
From the distribution of semi-major axis and eccentricity, we can determine the distribution of merging time for such binaries.
The merging time is given by \citep{1964PhRv..136.1224P}
\begin{equation}
T_{\rm mer} = \frac{5}{64}\frac{c^5a_0^4(1-e_0^2)^{7/2}}{G^3m_1m_2(m_1+m_2)}
\Bigg\{ 1+\frac{73}{24}e_0^2+\frac{37}{96}e_0^4 \Bigg\}^{-1}
\label{eqns21}
\end{equation}
where $a_0$ and $e_0$ are the initial semi-major axis and eccentricity, respectively.
Since the orbits of binaries in our consideration are nearly parabolic, the merging time has a strong dependence on
the velocity dispersion, i.e.,  $T_{\rm mer} \approx a_0^4(1-e_0)^{7/2} \approx \sigma^{-3}$.
Note that this merging time is not corrected for lower order PN terms (see \S 7 and Appendix A for other PN corrections).
\begin{figure}
  \centering
  \includegraphics[width=84mm]{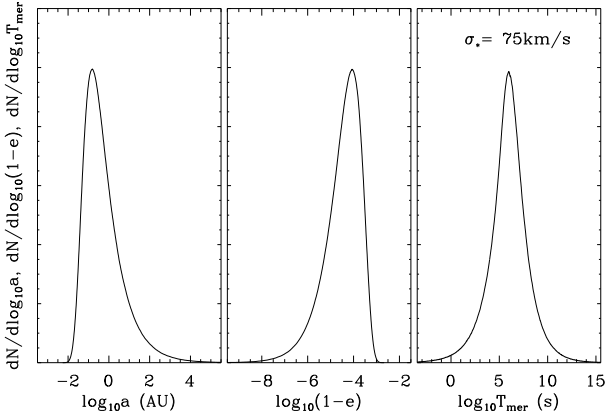}
  \caption[Distribution of semi-major axis, eccentricity and merging time]
  {Distribution of semi-major axis (left panel), eccentricity (middle panel) and merging time (right panel) 
  for $10M_{\odot}$ BH-BH binaries in the Milky-Way-like galaxies (i.e., $\sigma$ = 75 km/s).}
  \label{figg_dist}
\end{figure}
In Fig. \ref{figg_dist}, we have shown the distribution of semi-major axis, eccentricity and merging time of BH-BH binaries 
from equations (\ref{eqns19}), (\ref{eqns20}) and (\ref{eqns21}). As mentioned, we assume that the velocity of star follows 
one-dimensional normal distribution with $\sigma$= 75 km/s (i.e., the velocity dispersion for the Milky Way) and the pericentre distance $r_{\rm p}/r_{\rm p,max}$ follows uniform 
distribution. The peak position of each distribution is equivalent to the typical value in equations 
(\ref{eqns19}) and (\ref{eqns20}).
For merging time, the mode is $\sim10^{6}$ second.
Although the distribution is quite wide, almost all binaries will merge within a few thousand years.

\subsection{Merger rates}
\begin{figure}
  \centering
  \includegraphics[width=84mm]{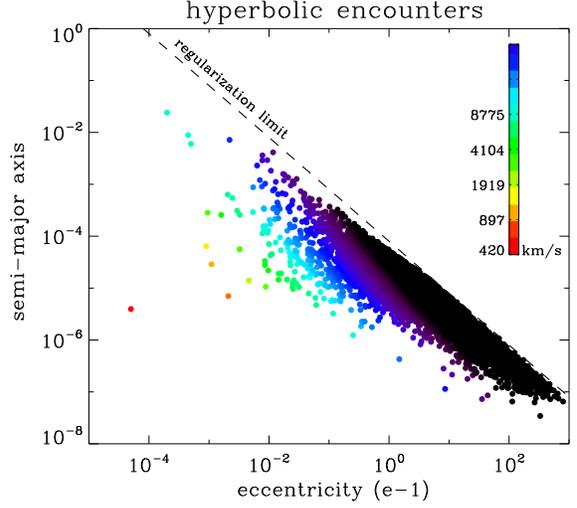}
  \caption[Semi-major axis and eccentricity of GR captured binaries]
  {Distribution of semi-major axis and eccentricity of GR captured binaries for the Model 4. 
  Different colors mean the velocity dispersion at the radius of influence in physical units as indicated by the color bar. 
  These encountering stars become binaries by GR capture.}
  \label{figg_mgr}
\end{figure}

We collect the parameters of all close encounter events in our simulations in order to investigate the GR capture and the compact binary coalescences.
In the \nbody6 code, the regularization algorithm helps the calculation of very close orbits
with high precision. If the separation or the time step of stars becomes smaller than certain criteria,
the stars are taken away from the main loop, and their motions are calculated with time smoothing.
We turn on the KS regularization, the two-body regularization scheme, and extract the semi-major axes 
and the eccentricities of pairs experiencing close encounters at the pericentre passage 
to avoid the effect of perturbation by nearby stars. We have shown an example of $(a,e)$ distribution of the 
close hyperbolic encounters for the Model 4 after $T>t_{\rm MBH}$ in Fig \ref{figg_mgr}.
Each filled dot represents each encounter and the dashed line shows the limit of KS regularization (i.e., $r_{\rm peri}\sim10^{-4}$), 
so pairs lying above this line are not considered in our investigation. 
In order to determine whether a certain encounter results in a binary, 
we therefore need to convert our dimensionless results to physical quantities according to the equation (\ref{eqns5}). 
In Fig. \ref{figg_mgr}, there are some orbits with rainbow colors. These colored orbits will become binaries
when the overall velocity dispersion of stellar system is larger than that velocity
(e.g., the orbit colored red at the left-end will become a binary when the velocity dispersion is $\sim$400 km/s.).
By counting the number of capture events, we estimate the binary formation rates in our simulations.

For a single BH passing through stars with a speed $v$, the time scale for GR capture is 
\begin{equation}
t_{\rm cap} \equiv  (\Sigma_{\rm cap}nv)^{-1}
\end{equation}
where $n$ is the number density of background stars. 
The binary formation rate between stars with different mass $m_1$ and $m_2$ 
in the shell with the range [$r$, $r+dr$] can be expressed as 
\begin{equation}
\frac{d\Gamma_{\rm cap}}{dr} = 4\pi r^2 n_1(r) n_2(r) \langle \Sigma_{\rm cap} v \rangle
\end{equation}
where $\langle \Sigma_{\rm cap} v \rangle$ is the velocity averaged value.
Thus, assuming $v_{\infty}=\sqrt{2}v$ in equation (\ref{eqns16}) and replacing $v$ by $\sigma(r)$, the velocity dispersion,
we have
\begin{equation}
\frac{d\Gamma_{\rm cap}}{dr} \simeq 87 \frac{G^2 m_1 m_2 \eta^{-5/7}}{c^{10/7}} r^2 n_1(r) n_2(r) \sigma(r)^{-11/7}.
\label{eqns24}
\end{equation}
For the case of systems composed identical stars, $m=m_1=m_2$, this equation becomes
\begin{equation}
\frac{d\Gamma_{\rm cap}}{dr} \simeq \frac{1}{2} 87 \frac{G^2 m^2 (1/4)^{-5/7}}{c^{10/7}} r^2 n^2(r)\sigma(r)^{-11/7}
\label{eqns25}
\end{equation}
where the half is to avoid double counting and $\eta=1/4$.
\begin{figure}
  \centering
  \includegraphics[width=84mm]{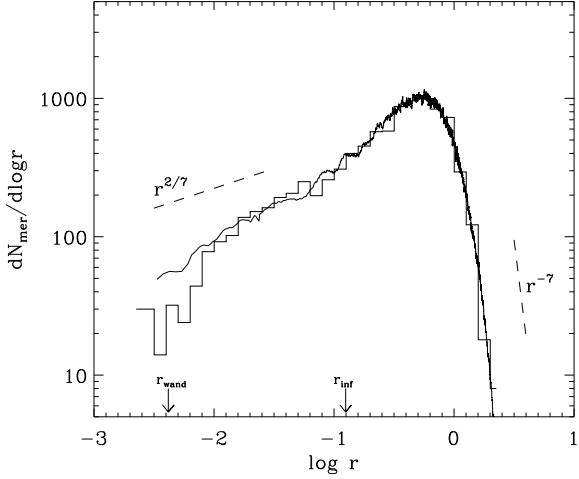}
  \caption[Radial distribution of GR binary capture rates]{Radial distribution of GR binary capture rates. 
  The solid curve is the semi-analytic model from the equation (\ref{eqns25}). This semi-analytic model agrees well with events in the simulation except
  for the inner region. From the theoretical models, the slope of distribution inside the radius of influence is expected to be 2/7.
  Although the actual slope differs from the theoretical models due to the overall incompleteness of Bahcall-Wolf cusp as shown in Fig. \ref{figg_gam}, the contribution to the overall merger rate within this radius is small. Thus, it does not significantly affect to the estimation of the entire merger rates.}
  \label{figg_dmgr}
\end{figure}
Fig. \ref{figg_dmgr} shows the radial distribution of cumulative capture events $dN_{\rm cap}/d\log r$ during $\Delta t = 2000 T_{\rm NBODY}$.
Histogram is from $N$-body simulations of Model 4, and the noisy line is from the equation (\ref{eqns25})
(i.e., $d N_{\rm cap}/d\log r = r d\Gamma_{\rm cap}/dr \cdot \Delta t$)
with the density and the velocity dispersion profiles from Figs. \ref{figg_rho} and \ref{figg_sig}, respectively.
In order to get large sample size, we set the unit of velocity to the half of the speed of light ($0.5c$). 
Although this velocity is unrealistically high, it is possible to guess what happens in realistic situations 
because there is no relativistic effect on the simulations.
The simulation results and our semi-analytic model from given density and velocity dispersion profiles show good agreement at the radius larger than $r_{\rm inf}$.
However, at $r<r_{\rm inf}$, there is some discrepancy between them: 
the binary formation rate obtained with simulation is smaller than that with the equation (\ref{eqns25}).
One possible reason is the time variation of density structure within the radius of influence as shown in Fig. \ref{figg_gam}.
From the theoretical model of star distribution within the radius of influence 
(e.g., $\rho(r) \sim r^{-7/4}$ and $\sigma(r)\sim r^{-1/2}$),
we obtain the slope of $d N_{\rm cap}/d\log r$ as
\begin{equation}
\frac{d N_{\rm cap}}{d\log r} \propto r^{2/7}.
\end{equation}
However, the slope inside $r_{\rm inf}$ in Fig. \ref{figg_dmgr} is not the same as $\frac{2}{7}$ 
because of the incompleteness of the Bahcall-Wolf cusp (see Fig. \ref{figg_gam}).
Around the wandering radius, even merger events from $N$-body simulations and the semi-analytic model show disagreement.
This gap is likely due to escapers that cause indirect heating at that radius. 
Incidentally, more than 80 per cent of events occurred outside $r_{\rm inf}$, 
and the peak of $d N_{\rm cap}/d\log r$ is located at $r_{\rm half}$. 
The fact that the most of the merger takes place outside the radius of influence is mainly because
the density distribution follows Bahcall-Wolf or shallower. Our results are quite different
from \cite{2009MNRAS.395.2127O} who found that the merger rate is more centrally concentrated
because the distribution of the black holes is much steeper than Bahcall-Wolf due to the mass segregation \citep[see Fig. 7 of][]{2009MNRAS.395.2127O}.
Our simulation is one component, and thus cannot reproduce the effect of the mass segregation.
We note here that many recent simulations with different approaches showed that 
the density profile of the massive component follows the Bahcall-Wolf or steeper while lower mass component
follows shallower slope \citep[direct $N$-body simulations,][]{2004ApJ...613.1143B} \citep[Monte-Carlo simulations,][]{2006ApJ...649...91F} and (time-dependent Fokker-Planck equation, Fiestas et al., in preparation). 
If that is the case, the small discrepancy of analytical estimation at small radii thus does not affect the estimation of the total capture rates.

In order to obtain the overall merger rate for NC, we need to integrate 
the equation (\ref{eqns25}) over the volume.
We can assume that the merger rate is equivalent to the capture rate 
because the merging time of BH-BH binary in our simulations is negligible compared to the cluster time scales as discussed before.
%(see \S \ref{sec_dis}, for more details).
It is difficult to estimate the merger rate because $n(r)$ and $\sigma(r)$ are not simple function of $r$. Assuming that
the velocity dispersion remains a constant over the entire cluster, and replacing the integral $\int n^2 r^2 dr$ by
$N {\widetilde n}$, we can write the integrated merger rate as

\begin{align}
\Gamma_{\rm mer} &\approx m^2 \cdot N \cdot \widetilde{n} \cdot \sigma_{*}^{-11/7} \nonumber\\
%&\approx& M^2 \cdot V^{-1} \cdot \sigma_{*}^{-11/7} \nonumber\\
%&\approx& M^2 \cdot (M^{-1}\sigma_{*}^2)^3 \cdot \sigma_{*}^{-11/7} \nonumber\\
&\sim M^{-1} \cdot \sigma_{*}^{31/7}
\label{eqns27}
\end{align}
where $\widetilde{n}, \sigma_{*}$ and $M$ are mean number density, 
the velocity dispersion of the system and total mass of the cluster, respectively. We have used the virial
theorem to get the last relationship.
We see that the event rate is inversely proportional to the total mass of the cluster with rather steep dependence on 
the velocity dispersion of the cluster.
\begin{figure}
  \centering
  \includegraphics[width=84mm]{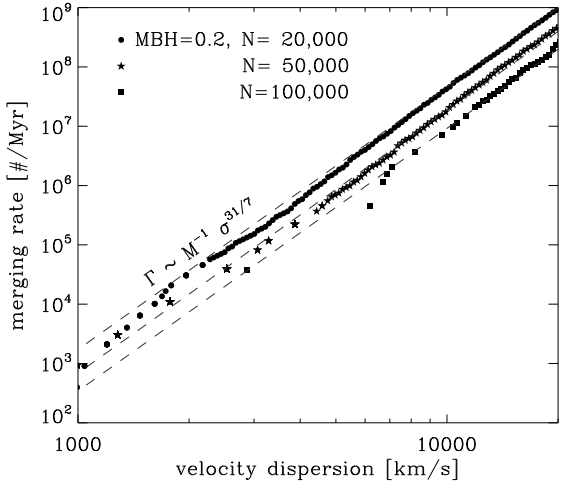}
  \caption[Merger rates as a function of velocity dispersion ($M_{\rm MBH}=0.2$)]
  {Merger rates as a function of velocity dispersion for models 2-4. The MBH mass is 20\% of the total mass of the cluster. Filled symbols are from the number counts 
  in simulations with different number of stars. 
  Dashed lines show the equation (\ref{eqns27}) with the proportional constants obtained from the simulations.
  There are good correlations between merger rates and velocity dispersion.}
  \label{figg_rat}
\end{figure}
\begin{figure}
  \centering
  \includegraphics[width=84mm]{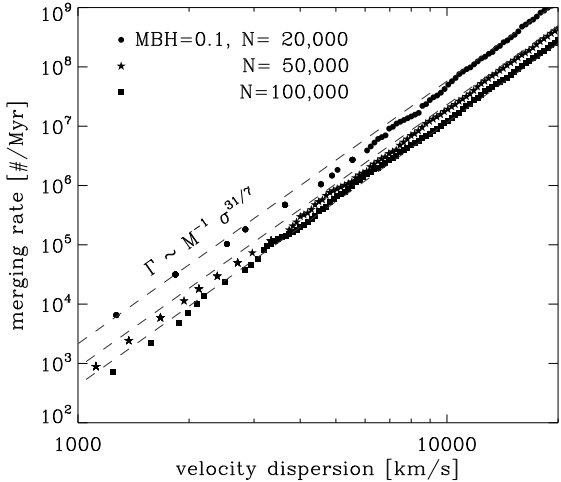}
  \caption[Merger rates as a function of velocity dispersion ($M_{\rm MBH}=0.1$)]
  {Merger rates as a function of velocity dispersion for models 5-7. The MBH mass is 10\% of the total mass of the cluster. }
  \label{figg_rat01}
\end{figure}
To convert our results to physical units, it is necessary to determine the representative value 
of velocity dispersion of $N$-body simulations.
We already estimated the density-weighted velocity dispersion in equation (\ref{eqns7}) as similar to the observational estimation of velocity dispersion of bulge as \citep{2013ApJ...764..184M}.
%\begin{equation}
%\sigma_{*}^2\equiv\frac{\int_{r_{\rm inf}}^{r_{\rm half}} 4\pi \sigma^2(r) \rho(r) r^2 dr}
%{\int_{r_{\rm inf}}^{r_{\rm half}} 4\pi \rho(r) r^2 dr},
%\end{equation}
%where $r_{\rm half}$ is half-mass radius. 
Note that the effective radius of bulge that is the upper limit of the integration for the systemic velocity dispersion in observations is much larger than the half-mass radius of NCs. According to recent observation for the calibration of velocity dispersion of nearby galaxies \citep{2013ApJ...767...26K}, however, the velocity dispersion does not change much with the aperture size. Thus, the velocity dispersion of NCs is  nearly identical to that of
bulges.

Figs. \ref{figg_rat} and \ref{figg_rat01} show the merger rates as a function of the velocity dispersion in the physical unit.
The mass ratio of MBH to the cluster is fixed at 0.2 for Fig. \ref{figg_rat} ($\sigma_* \sim 0.79$) and 0.1 for Fig. \ref{figg_rat01} ($\sigma_* \sim 0.75$), respectively.
Different symbols represent the different models with different $N$  and dashed lines are from 
the time averaged result of numerical integration of the equation (\ref{eqns27}).
Because of the limitations in the number of particles and the integration time, 
we only consider the unrealistic range of the velocity dispersion (1,000km/s $< \sigma_{*} <$ 20,000km/s).
Nevertheless, because the results show very good scaling relation with the velocity dispersion and the total mass of the cluster as predicted by the equation (\ref{eqns27}), 
it is possible to extrapolate our results to realistic parameters for NC.
As a result, the merger rate can be expressed by an equation with the mass of MBH and the velocity dispersion:
\begin{align}
\Gamma_{\rm mer}  \approx & 2.06 \times 10^{-4}{\rm Myr}^{-1} \Bigg( \frac{M_{\rm MBH}}{3.5 \times 10^6 M_{\odot}} \Bigg)^{-1} 
\Bigg( \frac{\sigma_{*}}{\rm 75 km/s} \Bigg)^{31/7}\nonumber \\
 & \textrm{ for } M_{\rm MBH} = 0.2 M_{\rm tot},
\label{eqns28}
\end{align}
\begin{align}
\Gamma_{\rm mer} \approx &  8.58 \times 10^{-5}{\rm Myr}^{-1} \Bigg( \frac{M_{\rm MBH}}{3.5 \times 10^6 M_{\odot}} \Bigg)^{-1} 
\Bigg( \frac{\sigma_{*}}{\rm 75 km/s} \Bigg)^{31/7}\nonumber \\
&  \textrm{ for } M_{\rm MBH} = 0.1 M_{\rm tot}.
\label{eqns29}
\end{align}
Thus, the merger rate is about 2.06$\times 10^{-10} {\rm yr}^{-1}$ for Milky-Way-like galaxies if we assume that the total mass 
of embedded star cluster is 5 times heavier than the MBH. 
\begin{figure}
  \centering
  \includegraphics[width=84mm]{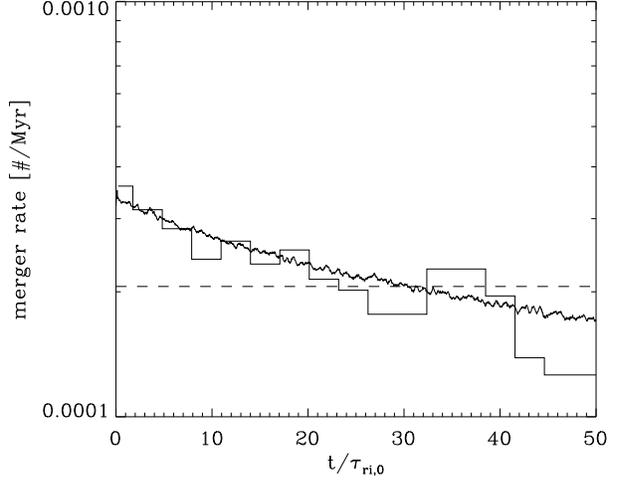}
  \caption[Time evolution of merger rate]
  {Time evolution of merger rate from Model 2 as a function of time at the radius of influence. The merger rate is scaled for a Milky-Way-like galaxy.
  Noisy line and histogram are from the integration of equation (\ref{eqns25}) and the number of events counted in 
  the simulations, respectively. Due to the cluster expansion, the merger rate decreases with time. The dashed horizontal line represents the merger rate from equation (\ref{eqns28}). }
  \label{figg_ratt}
\end{figure}

Our realization of the NC is not static: the NCs expand slowly with time even though they are bounded by the external
potential as described in \S4. The time evolution of the merger rate for a Milky-Way-like galaxy is represented in Fig. \ref{figg_ratt}, 
which is estimated from the Model 2 with $N=20,000$ in order to see the long-term evolution.
The time is scaled by the initial relaxation time at the radius of influence $\tau_{\rm ri,0}$ \citep{1987degc.book.....S} 
after growth of the MBH 
\begin{equation}
\tau_{\rm ri,0} \equiv \frac{\langle v^2\rangle^{3/2}}{15.4 G^2 \bar{m} \bar{\rho}_{\rm inf} \ln \Lambda}
\end{equation}
where $\bar{m}$ and $\bar{\rho}_{\rm inf}$ are the mean mass (for equal-mass, $\bar{m}=M/N$) 
and the mean density inside the radius of influence, respectively.
The noisy line is the merger rate from the numerical integration of the equation (\ref{eqns25}).
For comparison, we plot the histogram showing the number of events counted in the simulation.
They are selected with $\sigma_{*}=10^4$ km/s for sufficient samples and rescaled to $\sigma_{*}=75$ km/s range.
The horizontal dashed line is the time-averaged merger rate in equation (\ref{eqns28}).
Due to the expansion of the cluster, the merger rate decreases with time. These two estimates show good agreement, 
and therefore it is possible to surmise the merger rates from given density and velocity structures of stellar systems.
The conversion of these merger rates to the detection rates for GW detectors will be presented in next section.

\section{Detection Rates of BH Binary Mergers}
To determine the detection rate of GWs from BH-BH binary coalescences for GW detectors, 
it is necessary to calculate how many events occur per unit volume in the universe and 
horizon distance of GW detectors.
In the previous section, we estimated the merger rates in NC
as a function of the mass of MBH and the velocity dispersion.
It is well known that there is a good correlation between the mass of MBH 
and the velocity dispersion of surrounding stars \citep[e.g.,][]{2002ApJ...574..740T} 
\begin{equation}
M_{\rm MBH} \approx 1.3 \times 10^8 M_{\odot} (\sigma_*/200 {\rm km s^{-1}})^4
\end{equation}
in a range of the mass of MBH [$10^6 M_{\odot}$, $10^9 M_{\odot}$].
Later, \citet{2005ApJ...619L.151B} confirmed that the relation is also valid for MBHs down to $10^5 M_{\odot}$.
The merger rate for a NC, therefore, has a weak dependence on the mass of central MBH
$\Gamma \sim M_{\rm MBH}^{3/28}$ \citep{2009MNRAS.395.2127O}, and the merger rates of (\ref{eqns28}) and (\ref{eqns29}) become
\begin{align}
\Gamma_{\rm mer}&  \approx 3.33 \times 10^{-4}{\rm Myr}^{-1} \Bigg( \frac{M_{\rm MBH}}{3.5 \times 10^6 M_{\odot}} \Bigg)^{3/28} \nonumber \\
 &\quad \textrm{for } M_{\rm MBH} = 0.2 M_{\rm tot},
\label{eqns32}
\end{align}
\begin{align}
\Gamma_{\rm mer} &\approx 1.39 \times 10^{-4}{\rm Myr}^{-1} \Bigg( \frac{M_{\rm MBH}}{3.5 \times 10^6 M_{\odot}} \Bigg)^{3/28} \nonumber \\
& \quad \textrm{for } M_{\rm MBH} = 0.1 M_{\rm tot}.
\label{eqns33}
\end{align}

However, there are several factors that give rise to uncertainties in merger rates.
From the equation (\ref{eqns24}), we can infer that the merger rate is proportional 
to the total mass of different mass components $M_1$ and $M_2$.
As we mentioned before, however, we assumed that all stars are $10 M_{\odot}$ BHs. 
There exist other stellar objects such as MS stars, white dwarfs (WDs), NSs and BHs in real stellar systems.
\citet{2006ApJ...645L.133H} have studied the effect of mass segregation of stars around a MBH 
and concluded that the number fraction of different stellar objects evolves from the initial state 
\citep[i.e., $N_{\rm MS} : N_{\rm WD} : N_{\rm NS} : N_{\rm BH}$ = 1 : 0.1 : 0.01 : $10^{-3}$, for continuously star-forming populations;][]
{2005PhR...419...65A}
to $N_{\rm MS} : N_{\rm WD} : N_{\rm NS} : N_{\rm BH}$ = 1 : 0.09 : 0.012 : 0.06 within 0.1 parsec for Milky-Way-like galaxies.
If we set the mass of MS stars (0.7$M_{\odot}$), WDs (0.6$M_{\odot}$) and NSs 
(1.4$M_{\odot}$), the mass fraction of BHs $\mathcal{M}_{\rm BH}$ is about 44 per cent of the total mass.
When we simply assume that the merger rate of BHs in galactic nuclei can be expressed 
by the equations (\ref{eqns32}) and (\ref{eqns33}) with multiplication of $\mathcal{M}_{\rm BH}^2$,
the merger rate in equation (\ref{eqns32}) is reduced to 6.45$\times 10^{-11} {\rm yr}^{-1}$, 
which is about 3 times smaller than the estimation of \citet{2009MNRAS.395.2127O} for BHs with similar mass ranges.
Of course, it is more complicated to correct for the mass function rather than our consideration  because the mass fraction of BHs varies with the radius. Furthermore, the mass fraction in our consideration is adequate for innermost region although the capture events happen most frequently around the half-mass radii as shown in Fig. \ref{figg_dmgr}. The mass fraction of BHs around the half-mass radius might be smaller than that from \citet{2006ApJ...645L.133H}, and thus, our results could be an overestimation.

As discussed earlier, our model is one component, and the effect of the density profiles around the central
black hole may not be correct. If the density profile of the massive component is steeper significantly
for massive components, the merger rate could be enhanced \citep{2009MNRAS.395.2127O}. 
Moreover \citet{2009MNRAS.395.2127O} have shown that using different mass spectra of stellar mass BHs can affect not only the merger rates by a factor of $\sim$2 but also the detection rates by a factor of $\sim$10.
In that sense our result could be an underestimation. However, as we mentioned, Fiestas et al. (in preparation) found that the density profile of BHs does not significantly deviate from that of dominant components (low mass
stars) from time-dependent Fokker-Planck simulations.

The dynamical evolution of NCs also affects the merger rates.
The merger rate varies at most by a factor of $\sim$2 from $T=0$ to $T=200 \tau_{\rm ri,0}$ as shown in Fig. \ref{figg_ratt}.
\citet{2007ApJ...671...53M} have estimated the relaxation times $\tau_{\rm ri,0}$ 
for ACS Virgo samples of galaxies observed by \citet{2004ApJS..153..223C}, and found the relation between 
the relaxation time and the central velocity dispersion. According to the relation, 
$\tau_{\rm ri,0}$ is less than a Hubble time with smaller velocity dispersion than 100 km/s, 
corresponding to $M_{\rm MBH}\sim1.6\times 10^{7} M_{\odot}$. Therefore, the merger rates for NCs 
with smaller MBHs can be affected by the dynamical evolution.
In addition, the relaxation time of galactic nuclei implies that 
most of NCs with larger MBHs
do not contribute to the merger rates as much as those with smaller MBHs
because the number fraction of BHs in relaxed nuclei is several tens of times lager than that of 
initial conditions due to the mass segregation \citep{2006ApJ...645L.133H}, and
the merger rate weakly depends on the mass of central MBH \citep{2009MNRAS.395.2127O}.
\citet{2009MNRAS.395.2127O} also noted that the variance of the number density of galactic nuclei can affect the merger rate.
They have estimated the variance of the number density from the results of \citet{2007ApJ...671...53M} 
and found that the merger rate is enlarged as much with the rescale factor $\xi\sim 10-100$. However, see \citet{2013ApJ.777.203T}, for possible reduction of the rescale factor by about a factor of 5.

In order to calculate the merger rate per unit cosmological volume, we convolve the merger rate per NC
with the number density of MBHs in the universe \citep[for more details, see \S 3.3.5 of][]{2009MNRAS.395.2127O}.
\citet{2002AJ....124.3035A} determined the number density of MBHs from the luminosity function of galaxies as
\begin{equation}
\frac{dn_{\rm MBH}}{dM_{\rm MBH}} = c_{\bullet} \bigg(\frac{M_{\rm MBH}}{M_{\bullet}}\bigg)^{-\alpha}e^{-M_{\rm MBH}/M_{\bullet}}
\end{equation}
with the best fitting parameters of ($c_{\bullet}, M_{\bullet}, \alpha$) = 
($3.2\times10^{-11}M_{\odot}^{-1}{\rm Mpc}^{-3}, 1.3\times10^8M_{\odot},1.25$).
Note that using the MBH mass function of \citet{2002AJ....124.3035A} can lead to an overestimation of the detection rates at least by a factor of 2, especially at the low mass range of MBHs \citep{2007MNRAS.380L..15G}.
The merger rate per volume, therefore, is obtained by integrating the rate over the SMBH mass distribution
\begin{equation}
\mathcal{R}_{\rm mer}=\int^{M_{\rm u}}_{M_{\rm l}} \Gamma_{\rm mer}(M_{\rm MBH})\xi\frac{dn_{\rm MBH}}{dM_{\rm MBH}} dM_{\rm MBH}
\label{eqns35}
\end{equation}
where $M_{\rm u}$ and $M_{\rm l}$ are the upper and lower limits for integration, and $\Gamma_{\rm m,gal}$ and $\xi$ are the merger rate per galaxy as a function of the mass of MBH and the scale factor for number density variance, respectively.
The upper limit can be fixed to $M_{\rm MBH}\sim 10^{7} M_{\odot}$ by the time scale requirement as discussed above.
However, we still do not have the exact lower limit of the MBH mass,
which is currently about $10^5 M_{\odot}$ from the observation of \citet{2005ApJ...619L.151B}.
By taking the lower limit of the MBH mass to $10^4 M_{\odot}$, the equation (\ref{eqns35}) gives us the merger rate density
\begin{equation}
\mathcal{R}_{\rm mer}\approx 3.1 \Gamma_{\rm mer,MW}\xi_{30} {\rm Mpc}^{-3}
\label{eqns36}
\end{equation}
where $\Gamma_{\rm mer,MW}$ is the merger rate for a Milky-Way-like galaxy, and $\xi_{30}$ is the rescale factor for the variance of the number density of stars normalized by 30 \citep[i.e., $1/3 \le \xi_{30} \le 10/3$;][]{2009MNRAS.395.2127O}.
Choosing different lower limit of the MBH mass from $10^3 M_{\odot}$ to $10^5 M_{\odot}$ will give us an uncertainty by a factor of $\sim$3 in the merger rate density.

Now, we can estimate the detection rate of BH-BH binary coalescences by next generation GW detectors.
By assuming that the merger events occur uniformly in the universe, the detection rate only depends on the size of
cosmological volume which we can cover and can be expressed by \citep{2006ApJ...637..937O,2007ApJ...662..504B,2011MNRAS.416..133D, 2013MNRAS.440.2714}
\begin{equation}
\mathcal{R}_{\rm det} = \mathcal{R}_{\rm mer} \int \frac{4\pi r(z)^2}{1+z}\frac{dr}{dz}dz
\end{equation}
where $z$ is the cosmological redshift, and the factor of $(1+z)^{-1}$ represents the cosmological time dilation. 
For existing GW detectors, the effect of redshift can be negligible because their coverage is not too far 
\citep[i.e., the horizon distance $D_{\rm h}$ are 33 Mpc for NS-NS binaries and 161 Mpc for BH-BH binaries 
corresponding to $z \sim$ 0.01 and 0.04 in standard $\Lambda$CDM cosmology, respectively;][]{2010CQGra..27q3001A}.
However, for next generation GW detectors, the effect of redshift becomes important, especially for BH-BH binaries.
The maximum horizon distance $D_{\rm h}$ can be obtained from signal-to-noise ratio (SNR) of GW signals 
\citep[for more details, see \S 4.2 of][]{2009MNRAS.395.2127O}. 
Because the redshift affects both the mass of source and the frequency, SNR should be estimated from the waveforms carefully. 
Only few studies \citep{2007PhRvD..75l4024B,2009MNRAS.395.2127O,2009PhRvD..80l4026R} have estimated $D_{\rm h}$ for BH-BH binaries for given SNR.
\begin{table*}
  \begin{center}
  \caption{Detection rates of BH-BH binaries for advanced LIGO.}
  \begin{tabular}{l c c c c c c c}
    \\
    \hline
    \hline
    Models & $M_{\rm MBH}/M_{\rm cl}$ & $\Gamma_{\rm mer,MW}$ & $\mathcal{M}_{\rm BH}^a$ & $D_{\rm h}$ 
  & $\mathcal{R}_{\rm det,l}^b$ & $\mathcal{R}_{\rm det,re}^c$ & $\mathcal{R}_{\rm det,h}^d$ \\
  &   &   {\footnotesize (Myr$^{-1}$)} & & {\footnotesize (Mpc)} & {\footnotesize (yr$^{-1}$)} & {\footnotesize (yr$^{-1}$)} 
  & {\footnotesize (yr$^{-1}$)} \\
    \hline
    & & & & 986$^e$                                        & 0.05 &  0.39  & 3.0\\
    1-4 & 0.2 & $3.3\times10^{-4}$ & 0.44 & $\sim$1100$^f$ & 0.07 & 0.51 & 4.0 \\
    & & & & $\sim$1900$^g$                                 &  0.23 & 1.8 & 14\\
    \hline
    & & & & 986$^e$ & 0.02 & 0.16 & 1.3\\
    5-7 & 0.1 & $1.4\times10^{-4}$ & 0.44 & $\sim$1100$^f$ & 0.03 & 0.21 & 1.7 \\
    & & & & $\sim$1900$^g$ & 0.10 & 0.74 & 5.7 \\
    \hline
    \label{tblg2}
  \end{tabular}
  \end{center}
  \begin{flushleft}
   $^{a}$\small Mean mass fraction of BHs from \citet{2006ApJ...645L.133H}.\\
   $^{b,c,d}$\small Low, realistic and high detection rates depending on uncertainties.\\% discussed in text.\\
   $^{e}$\small$D_{\rm h}$ with SNR 8 form \citet{2010CQGra..27q3001A} divided by 2.26, 
   the correction factor for sky\\ location and orientation of sources.
   The effect of redshift is not included.\\
   $^{f}$\small$D_{\rm h}$ with SNR 10 form Fig. 16 in \citet{2007PhRvD..75l4024B}.\\
   $^{g}$\small$D_{\rm h}$ with SNR 8 form Fig. 3 in \citet{2009PhRvD..80l4026R}.\\
  \end{flushleft}
\end{table*}

Table \ref{tblg2} shows the detection rates expected for advanced LIGO. 
Because we mainly considered the equal-mass models for only BHs, the detection rates are corrected by the factor of $\mathcal{M}_{\rm BH}^2$.
We estimated detection rates by using $D_{\rm h}$ from different studies
\citep{2007PhRvD..75l4024B,2009PhRvD..80l4026R,2010CQGra..27q3001A}. 
All $D_{\rm h}$s are corrected for the orientation of sources. Note that $D_{\rm h}$ from \citet{2010CQGra..27q3001A}
does not include the cosmological effect of the redshift. $D_{\rm h}$ from \citet{2009PhRvD..80l4026R} is for spinning BHs but 
 independent of the spin of BHs for 10$M_{\odot}$ BHs.
We list three detection rates $\mathcal{R}_{\rm det,l}$, $\mathcal{R}_{\rm det,re}$ and $\mathcal{R}_{\rm det,h}$ in Table \ref{tblg2} for different models and different estimates of detection horizon ($D_{\rm h}$). 
We first calculated the `reasonable' expected detection rate $\mathcal{R}_{\rm det,re}$ using the merger rate given in
equation (\ref{eqns36}). Given several uncertainties including the variance of the number density, dynamical evolution
and the lower limits of the MBH, we assume that the total range of uncertainty is a factor of 60. Thus `low' expected
rate  $\mathcal{R}_{\rm det,l}$  is simply obtained by dividing  $\mathcal{R}_{\rm det,re}$ by  $\sqrt{60}$ while the
`high' expected rate $\mathcal{R}_{\rm det,h}$ is calculated by multiplying the same factor. The whole range of
the expected detection rate lies between 
 0.02 $\sim$ 14 yr$^{-1}$ depending on the maximum horizon distance and uncertainties.
These estimates are able to cover those of \citet{2009MNRAS.395.2127O} (5-20 yr$^{-1}$) for BHs around $10 M_{\odot}$ although the lowest value is quite smaller than that.

Our estimations have some limitations;
(1) We ignore the initial mass function. This mass function may affect not only the evolution of systems by the relaxation 
between mass components but also the merger rates for BHs with different masses.
(2) We need to consider various range of $M_{\rm MBH}/M_{\rm cl}$ 
\citep[e.g.,][ suggested that there is a rough relation between the mass of MBH and NC.]{2009MNRAS.397.2148G}
(3) The exact calculation for SNR is necessary in order to obtain more reasonable detection rates.  We considered
the mass segregation and its effect only approximately. 
These limitations would be considered in future works.

\section{Black hole binary coalescence and waveform}
Solving Einstein field equation exactly is very difficult and has only been done numerically. 
Fortunately, during inspiral phase of compact binary coalescences, the Einstein equation can be simplified 
with the PN expansion. 
When a compact binary is formed, the orbit decays with time due to the GR.
%overview%
The orbit-averaged change of the semi-major axis and the eccentricity by GR is first derived by 
\citet{1964PhRv..136.1224P} as 
\begin{equation}
\bigg< \frac{da}{dt} \bigg> = -\frac{64}{5} \frac{G^3m_1m_2(m_1+m_2)}{c^5 a^3 (1-e^2)^{7/2}} 
\bigg( 1+ \frac{73}{24}e^2 + \frac{37}{96}e^4 \bigg),
\label{eqns38} 
\end{equation}
\begin{equation}
\bigg< \frac{de}{dt} \bigg> = -\frac{304}{15} e \frac{G^3m_1m_2(m_1+m_2)}{c^5 a^4 (1-e^2)^{5/2}}
\bigg( 1+ \frac{121}{304}e^2 \bigg)
\label{eqns39}
\end{equation}
in the 2.5PN order ($\sim 1/c^5$, the first order GR term).
However, the orbital evolution is also affected by other PN order terms such as 1PN (relativistic precession),
1.5PN (spin-orbit coupling), 2PN (spin-spin coupling, high order relativistic precession) and higher orders.
With full consideration of PN terms up to 2.5 order, \citet{2009ApJ...695..455B} noted that the decay of the binary orbit is much faster than that with 2.5PN only.
Although the effect of the spin is quite important for the motions and waveforms of BH-BH binary coalescences, 
we only consider non-spinning BHs and take 1, 2 and 2.5PN order terms in this study.

%In order to investigate the orbital evolution of compact binaries in general relativity, 
%the exact equation of motion with full PN corrections is needed. 
The equation of motion with PN correction up to 2.5PN order can be simply written in the centre of mass frame as
\citep{2003CQGra..20..755B,2004PhRvD..69j4021M}
\begin{figure}
  \centering
  \includegraphics[width=84mm]{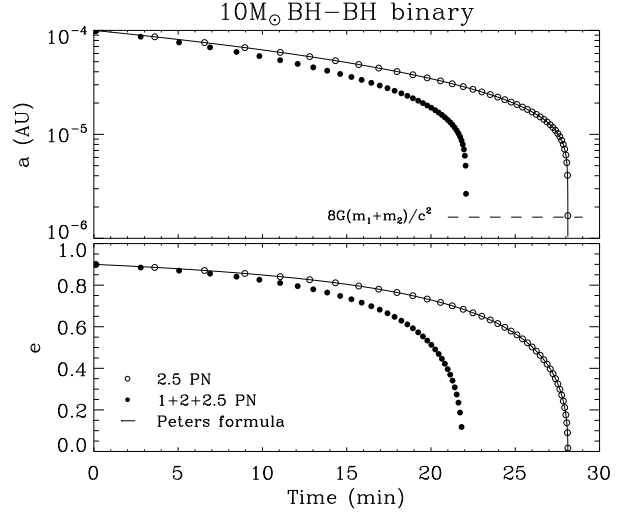}
  \caption[Orbital evolution of coalescing binary]
  {Time evolution of semi-major axis and eccentricity of a coalescing binary. Solid line is from the integration of
  Peters formula. Open and filled circles are the results of simulations with 2.5PN correction only and full PN
  correction, respectively. The simulation with 2.5PN correction only agrees well with Peters formula
  while that with full PN correction significantly festinates compared to Peters formula.}
  \label{figg_orb0}
\end{figure}
\begin{equation}
\mathbf{a} = \mathbf{a_{n}} + \mathbf{a_{pn}} = - \frac{Gm}{r^3}\mathbf{r} + \frac{Gm}{r^2}(A\frac{\mathbf{r}}{r}+B\mathbf{v})
\end{equation}
where $A, B$ are PN coefficients depending on their masses, the relative position $r$ and relative velocity 
$\mathbf{v}$ (see Appendix A, for more details).
Many authors have incorporated the PN corrected force in direct $N$-body simulations with different PN orders 
\citep{1993ApJ...418..147L,2007MNRAS.378..285A,2009ApJ...695..455B,2014MNRAS.437.1259B}.
Similarly, we implemented the PN equation of motion to the KS regularization process in \nbody6 code.
In the KS regularization process, a two-body motion is sometimes perturbed by other neighboring stars, 
and these perturbation should be corrected. Thus, we can consider the PN force as a perturbing force in the code
by adding the PN force and its time derivative. We designed that the binary will merge 
when the separation is smaller than four Schwarzschild radii $4R_{\rm Sch} \equiv 4\cdot2G(m_1+m_2)/c^2$ 
because the PN approximation is not valid in this regime any more.

For instance, Fig. \ref{figg_orb0} shows the time evolution of semi-major axis and eccentricity 
for a 10$M_{\odot}$ BH-BH binary with PN approximation. 
The initial semi major axis and eccentricity are $10^{-4}$ AU and 0.9, respectively. 
The solid line represents the time integration of Peters formula, equations (\ref{eqns38}) and (\ref{eqns39}).
There is a good agreement between the results of simulation with only 2.5PN term (open circle) 
and integration of Peters formula.
On the other hand, the merging time of simulation with all terms up to 2.5PN (filled circle) 
is significantly smaller than that of Peters formula as reported in \citet{2009ApJ...695..455B}.
In case of this binary, it takes less then a half hour for merging (i.e., $r_{12} \le 4R_{\rm Sch}$).

In our simulations, most of pairs of close encounters have not been perturbed by other nearby stars. 
Thus, it is possible to separate the two-body motion with PN correction from the main loop of simulations.
We, here, use a \textsc{toy}\footnote{http://www.ast.cam.ac.uk/{\footnotesize$\sim$}sverre/web/pages/nbody.htm} 
code only for a KS two-body motion written by S. J. Aarseth
for convenience of exploration of the evolution of merging binaries.
The PN implementation mentioned earlier is also adopted in the \textsc{toy} code.
For given semi-major axis and eccentricity, we simulate the orbital evolution of binaries.
\begin{figure}
  \centering
  \includegraphics[width=84mm]{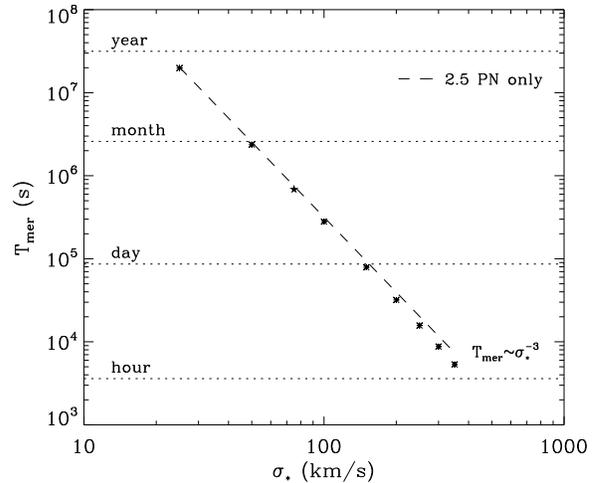}
  \caption[Merging time of BH-BH binaries with different velocity dispersion]{Merging time of typical 10$M_{\odot}$ (see \S\ref{subsec5.1} for details)
  BH-BH binaries with different velocity dispersion of embedded star clusters. The merging times are obtained 
  from two-body simulations with all corrections up to 2.5PN. The range of velocity dispersion is 50 to 400 km/s, which is correspond to
  the range of the mass of SMBH from $5\times 10^5 M_{\odot}$ to $2\times 10^9 M_{\odot}$ according to 
  the $M_{\rm MBH}-\sigma_{*}$ relation from \citet{2002ApJ...574..740T}. 
  Star symbol represents the binary merging time in the Milky-Way-like galaxies. In all cases, the merging times are smaller than a year.}
  \label{figg_tmr}
\end{figure}
Fig. \ref{figg_tmr} shows the merging time of typical BH-BH binaries in equations (\ref{eqns19}) and (\ref{eqns20}) 
with different velocity dispersion of systems. The merging time is estimated by two-body simulations with all PN corrections.
A star symbol is showing the result of a galactic nucleus in Milky-Way-like galaxy.
In all cases, the merging times are less than a year. Binaries formed by GR capture, therefore, will merge immediately.

The GW waveforms of coalescing binaries have been studied by many authors 
\citep[e.g.,][]{1990PhRvD..42.1123L,1995PhRvD..52..821K}.
As the perturbation of flat-space metric, $h^{ij}$ in can be expressed by 
\citep[for more details, see equations 3.21 and 3.22 in][]{1995PhRvD..52..821K}
\begin{align}
h^{ij} &= \frac{2\mu}{D}\bigg[ Q^{ij} + P^{0.5}Q^{ij}+P(Q^{ij}+Q^{ij}_{\rm SO})\nonumber \\
&+P^{1.5}(Q^{ij}+Q^{ij}_{\rm SO})+P^2Q^{ij}_{\rm SS}+\cdots \bigg]_{\rm TT}
\end{align}
%in the unit of $G=c=1$, 
where $\mu$ is the reduced mass, $D$ is the distance from the source to the detector, 
$Q^{ij}$ is the time derivative of quadrupole moment tensor, $P^n$ is the PN corrections with order of $n$, 
and SO, SS and TT denote spin-orbit coupling, spin-spin coupling and transverse-traceless gauge, respectively. Here $G=c=1$ is used.
Since we are interested in the aspects of GW rather than the exact waveforms, we take the leading order of $h^{ij}$ 
\begin{equation}
h^{ij}\approx\frac{4\mu}{D}\Bigg[v^iv^j-\frac{m}{r}n^in^j\Bigg]
\end{equation}
with 
\begin{equation}
Q^{ij}=2\Bigg[v^iv^j-\frac{m}{r}n^in^j\Bigg]
\end{equation}
where $v^i$ and $n^i$ are the relative velocity and the normal vector of the relative position, respectively.
It is well known that GWs have two polarization $+$ and $\times$ and 
waveforms are the linear combination of these two polarizations.
If we assume that the orbital plane lies on the {\it xy} plane initially in the source coordinate, 
and the angle between the direction to the detector $\hat{\textbf{N}}$ and the angular momentum $\hat{\textbf{J}}$ is $\Theta$, 
the polarizations $h_+$ and $h_\times$ are given by \citep{1995PhRvD..52..821K}
\begin{equation}
h_+ = \frac{1}{2} \bigg( \cos^2\Theta h^{xx} - h^{yy} +\sin^2\Theta h^{zz} -\sin2\Theta h^{xz} \bigg),
\end{equation}
\begin{equation}
h_\times = \cos\Theta h^{xy} - \sin\Theta h^{yz}.
\end{equation}

Now we provide a waveform of a typical 10$M_{\odot}$ BH-BH binary coalescence in a Milky-Way-like galaxy for an example.
The semi-major axis and eccentricity after GR capture are 0.153 AU and 0.99989 from the equations 
(\ref{eqns19}) and (\ref{eqns20}), respectively.
\begin{figure}
  \centering
  \includegraphics[width=84mm]{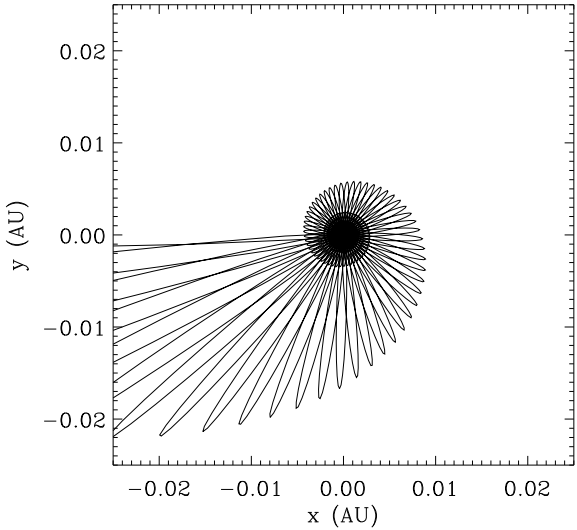}
  \caption[Orbital motion of coalescing 10$M_{\odot}$ BH-BH binary]{Relative orbital motion of a BH-BH binary with 
  the initial semi-major axis 0.153 AU and eccentricity 0.99989, as a representative of typical BH-BH binaries 
  in Milky-Way-like galaxies. The orbits are very eccentric at the beginning. 
  The perihelion is shifted counterclockwise due to the 1PN and 2PN terms, and the orbit shrinks with time due to the GW emission.}
  \label{figg_xy}
\end{figure}
In Fig. \ref{figg_xy}, the relative motion of BHs on {\it xy} plane is shown. 
Due to the 1PN and 2PN terms, the position of perihelion is shifted counterclockwise.
In addition, by emitting GWs, the orbit shrinks more and more with time.
\begin{figure*}
  \centering
  \includegraphics[width=170mm]{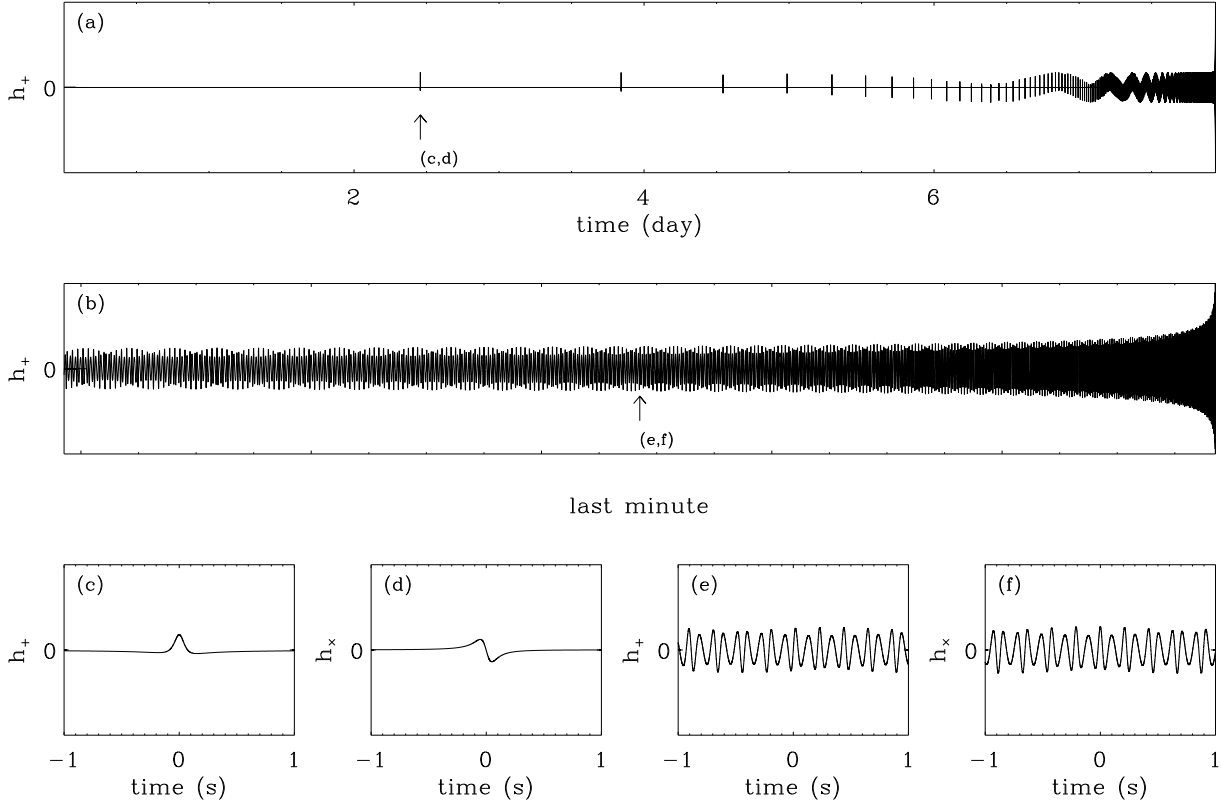}
  \caption[Waveform of a BH-BH binary coalescence]{Waveform of BH-BH binary coalescence  for the same binary in Fig \ref{figg_xy}.
  (a) $h_+$ for whole stage. The merging time is $\sim$8 days. The waveform is like a burst, initially. 
  (b) $h_+$ for last minute. The waveform is much sinusoidal at this time. 
  (c,d) $h_+$ and $h_\times$ for the first burst as marked in (a). 
  They are similar to those of eccentric binaries in \citet{1992Sci...256..325A}. 
  (e,f) $h_+$ and $h_\times$ at a half minute before coalescence as marked in (b). 
  At this time, the frequency is $\sim$ 10 Hz.}
  \label{figg_wav}
\end{figure*}
Fig. \ref{figg_wav} shows the waveforms for this BH-BH binary coalescence. 
For simplicity, we assume that the axis of angular momentum is aligned with the direction to the detector 
(i.e., face-on view, $\Theta=0$). 
In Fig. \ref{figg_wav}(a), the waveform of $+$ polarization during whole evolution is presented. 
The merging time is about 8 days.
Interestingly, the waveform is burst-like at the beginning, and it takes more than 2 days for the first burst after capture.
The detailed waveforms $h_+$ and $h_\times$ at this moment are shown in \ref{figg_wav}(c) and (d). 
These waveforms are similar to those of eccentric orbits in \citet{1992Sci...256..325A}. 
Fig. \ref{figg_wav}(b), (e) and (f) show the waveform in the last minute, 
the detailed view of $h_+$ and $h_\times$ a half minute before merging, respectively. 
In this stage, the orbit is much circularized compared to the beginning, and the orbital frequency is about 10 Hz.
At the moment of coalescence, the orbital frequency becomes few hundreds Hz 
which is the detectable frequency by ground-base GW detectors.

\begin{figure}
  \centering
  \includegraphics[width=84mm]{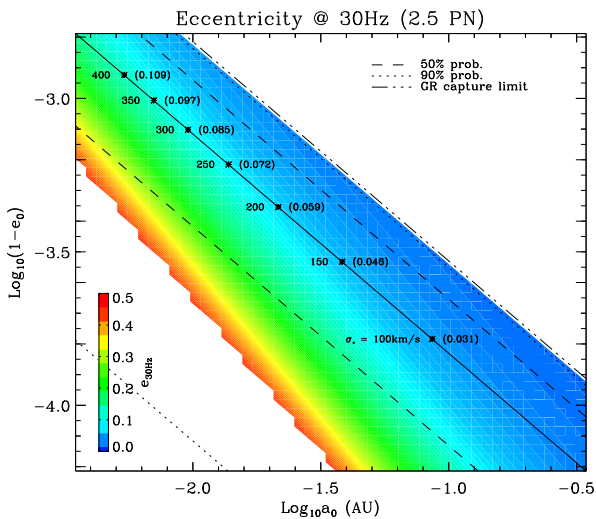}
  \caption[]{Eccentricity of BH binaries at 30 Hz which is the typical low frequency limit of ground based
detectors, in the initial ($a_{0},e_{0}$) plane. The eccentricity at 30 Hz is expressed by the color and the diagonal
solid line represents the typical values of $(a_{0},e_{0})$ for different $\sigma_*$.  We limited the range of calculation
down to 10\% of the probability peak of the initial probability distribution function of the initial eccentricity for
a given velocity dispersion marked at the central solid line. Also we show the locus of 50 \% of the peak as broken
lines.}

  \label{figg_ecc}
\end{figure}

One of the important questions with regards to the GR captured binaries is the typical value of eccentricity
when they enter the detectable band of the ground-based detectors. We computed the eccentricity when GW frequency 
becomes 30 Hz as a function of initial orbital parameters as shown in Fig. \ref{figg_ecc} using the post-Newtonian
approximation with 2.5PN term. The diagonal line is the locus of
the typical   $(a_0, e_0)$ plane for different velocity dispersion, and we computed the eccentricities at 30 Hz for
up to 10\% of peak in probability distribution function along the initial eccentricity for a given velocity dispersion.
Different colors represent different
eccentricity at 30 Hz. As can be seen from this figure, the GR captured binaries from typical NC with $\sigma_* \approx 100$
km/sec become nearly circular when they enter the LIGO/Virgo band. Only those from
very massive NC with high velocity dispersion are likely to remain eccentric up to the high frequency band. Note that we have
computed the orbital evolution only up to the pericentre distance of 4 $R_{\rm Sch}$ and some extreme binaries
will not emit GW signal higher than 30 Hz. This could be an artifact of using post-Newtonian approximation. Eventually
the binaries in this regime should emit gravitational waves at higher frequencies than 30 Hz. We simply
refer to \citet{2013PhRvD..87d3004E} for the derivation of more realistic derivation  of the waveforms for eccentric
binaries and their detectability with the ground-based GW detectors.

\section{Summary and Conclusion}

We have generated $N$-body realizations for nuclear star clusters (NCs) located at the centre of galactic bulges and hosting a massive black hole (MBH). In our simulations, the surrounding bulge is considered as the external potential well which makes the velocity dispersion of the embedded star cluster
isothermal since a deep potential well behaves like a heat bath \citep{2011MNRAS.414.2728Y}.
The MBH, in the same manner, is also modeled as a point-mass potential but growing with time to ensure
the adiabatic adjustment of the stellar system. Consequently, our $N$-body realizations have a stellar
density cusp \citep[$ \rho \sim r^{-1.75}$;][]{1976ApJ...209..214B} and Keplerian velocity
dispersion within the radius of influence. In addition, the overall velocity structure is similar
to observations of the star cluster at the centre of Milky Way \citep[e.g.,][]{2009A&A...502...91S}. Strictly speaking, however, these star clusters are not in equilibrium but
expand continuously since the MBH can generate kinetic energies by the interaction with stars in the cusp. The slopes of density and velocity dispersion profiles of our $N$-body realizations are slightly shallower than those of theoretical expectations.

This environment of NCs is a good laboratory for gravitational wave (GW) sources. In order to investigate GW event rates in NCs, we have collected
the orbital information of close encounters (i.e., semi-major axis and eccentricity) in our $N$-body simulations.
While most of binaries are disrupted by the strong tidal field from MBH, there can be many hyperbolic encounters of stellar mass black holes (BHs) whose pericentre distances are sufficiently small to radiate GWs efficiently due to the high density and velocity dispersion at the vicinity of the MBH and the high number fraction of BHs due to the mass segregation \citep{2006ApJ...645L.133H,2009MNRAS.395.2127O}.
When the energy loss by gravitational radiation (GR) is greater than the orbital energy, two BHs make a binary and merge
 quickly because of the small separation and large eccentricity after capture. Thus, the capture event rate corresponds to the merger rate.
The capture happens most frequently near the  half mass radius rather than within the radius of influence. Thus, our investigation of GR capture event rates is still valid although our models can not precisely realize the cluster inside of the radius of influence.
By counting the number of GR capture events, we have built scaling relations of merger rates for a NC as a function of the mass of the MBH and the velocity dispersion of the star cluster. As the result, the merger rate for a Milky-Way-like galaxy is $\sim 10^{-10}{\rm yr}^{-1}$ proportional to the mass ratio of MBH to the star cluster.

From the $M_{\rm MBH}-\sigma_{*}$ relation \citep[e.g.,][]{2002ApJ...574..740T}, the merger rate becomes a function of the mass of MBH only. By using realistic mass function of MBHs \citep[e.g.,][]{2002AJ....124.3035A}, we have determined the merger rate density per unit volume.
Then, the detection rates can be expressed with the merger rate density and the size of cosmological volume covered by GW detectors if we simply assume the uniformness of merger events over cosmic time and volume.
 We have obtained the expected detection rates $0.02 \sim 14 {\rm yr}^{-1}$ for advanced LIGO depending on
the maximum horizon distances from different studies \citep{2007PhRvD..75l4024B,2009PhRvD..80l4026R,2010CQGra..27q3001A} and the mass ratio of MBH to the star cluster 0.1 and 0.2.
%The numbers in parentheses refer to the corrected values of following uncertainties in our estimation: the dynamical
% evolution of the cluster (by a factor of $2\sim3$), the variance of number density of stars (by a factor of $\sim$10) and 
%the mass range of MBH (by a factor of $2\sim3$).}
This estimate can cover those of \citet{2009MNRAS.395.2127O} who suggested the detection rate of $\sim 5-20{\rm yr}^{-1}$ of 5 to 15 $M_{\odot}$ BH-BH binary coalescences in galactic nuclei for advanced LIGO.
However, this study still have some limitations; (1) It is necessary to consider realistic mass function for BHs instead of assuming the mean mass fraction of BHs. (2) There is a relation between the mass of MBHs and the NCs \citep{2009MNRAS.397.2148G}. However, we fixed the mass ratio of MBH to the star cluster to 0.1 and 0.2. (3) In order to determine the maximum horizon distance for BH-BH binary mergers, the precise signal-to-noise ratio calculation is needed.

We have investigated the statistics of coalescing BH-BH binaries and found that the typical semi-major axis and eccentricity of these binaries are related to the velocity dispersion of the system. We also have implemented the post-Newtonian (PN) approximation on the two body motions up to 2.5PN orders. With a given set of semi-major axis and eccentricity, we calculated the two-body motion under the PN approximation and the waveform of GW emission. The merging time is about a few hours for a typical BH-BH binaries in a Milky-Way-like galaxy.  We also estimated the eccentricities of GR captured binaries when they
enter the LIGO/Virgo band (assumed to be 30Hz). Most of the binaries formed in relatively low velocity environment
($\sigma_* \sim$ 100 km/s) would become nearly circular while those formed in very high velocity environment 
are likely to keep significant eccentricity when they cross the 30 Hz. Considering the fact that the low velocity 
galaxies contribute more to the total merger rate (cf. equation \ref{eqns35}), the GR captured binaries are likely to be
indistinguishable from those of different origin in the sense the eccentricity effect is nearly negligible.

\section*{acknowledgments}
This work was supported by the National Research Foundation of
Korea (NRF) grant NRF-2006-0093852 funded by the Korean government.
HML acknowledges Alexander von Humboldt Foundation
for the Research Award in 2013.

\section*{Appendix A. Post Newtonian Equation of Motion in Center of Mass Frame\label{app}}
%=============================================================================

Here, we present the post Newtonian (PN) equation of motion of binary system
in the centre of mass frame up to 2.5PN order by following \citet{2004PhRvD..69j4021M} \citep[also see Appendix in][]{2014MNRAS.437.1259B}.
Because we assume non-spinning BHs, we do not consider spin terms in this paper.
For the beginning, we borrow notations from \citet{2004PhRvD..69j4021M} as
\begin{align}
m&= m_1 + m_2, \nonumber \\
\mathbf{v}&= \mathbf{v}_2 - \mathbf{v}_1, \nonumber \\
\mathbf{r}&= \mathbf{r}_2 - \mathbf{r}_1, \nonumber \\
\mathbf{n}&= \mathbf{r}/r, \nonumber \\
\eta&= (m_1m_2)/(m_1+m_2)^2
\end{align}
where $m_1$ and $m_2$ are the mass of stars, $\mathbf{r}_1, \mathbf{r}_2, \mathbf{v}_1$ and
$\mathbf{v}_2$ are the 3-dimensional positions and velocities, and $\eta$ is the symmetric mass ratio.
As mentioned in the text, the PN acceleration can be considered as a perturbation
and added to the gravitational acceleration as
\begin{equation}
\mathbf{a} = \mathbf{a_{n}} + \mathbf{a_{pn}} = - \frac{m}{r^2}\mathbf{n} + \frac{m}{r^2}(A\mathbf{n}+B\mathbf{v})
\end{equation}
where $A$ and $B$ are the PN coefficients related to the relative position and velocity,
respectively,
where
\begin{equation}
A_1 = 2(2+\eta)\frac{m}{r}-(1+3\eta)v^2+\frac{3}{2}\eta\dot{r}^2,
\end{equation}
\begin{flalign}
A_2&= -\frac{3}{4}(12+29\eta)\frac{m^2}{r^2}-\eta(3-4\eta)v^4-\frac{15}{8}\eta(1-3\eta)\dot{r}^4 \nonumber\\
   &+\frac{1}{2}\eta(13-4\eta)\frac{m}{r}v^2+(2+25\eta+2\eta^2)\frac{m}{r}\dot{r}^2 \nonumber\\
   &+\frac{3}{2}\eta(3-4\eta)v^2\dot{r}^2,
\end{flalign}
\begin{equation}
A_{5/2} = \frac{8}{5}\eta\frac{m}{r}\dot{r}\bigg(\frac{17}{3}\frac{m}{r}+3v^2\bigg)
\end{equation}
and
\begin{equation}
B_1 = 2(2-\eta)\dot{r},
\end{equation}
\begin{equation}
B_2 = -\frac{1}{2}(4+41\eta+8\eta^2)\frac{m}{r}\dot{r}+\frac{1}{2}\eta(15+4\eta)v^2\dot{r}-\frac{3}{2}\eta(3+2\eta)\dot{r}^3,
\end{equation}
\begin{equation}
B_{5/2} = -\frac{8}{5}\eta\frac{m}{r}\bigg(3\frac{m}{r}+v^2\bigg)
\end{equation}
where $\dot{r}$ is the first time derivative of radial distance defined as
$\dot{r} = \mathbf{r}\cdot\mathbf{v}/r$.
Then the coefficients $A$ and $B$ are given by the summations of coefficients for different PN order
divided by the speed of light $c$, $A_{i}/c^{2i}$.
Note that the sign of $B$ is opposite of that in \citet{2003CQGra..20..755B}.

Since the \textsc{nbody }code uses 4th-order Hermite integrator, we have to have the first time derivatives of accelerations as similar to the accelerations
\begin{equation}
\mathbf{\dot{a}} = \mathbf{\dot{a}_{n}} + \mathbf{\dot{a}_{pn}}.
\end{equation}
The derivative of PN acceleration $\mathbf{\dot{a}_{pn}}$ can be expressed as
\begin{equation}
\mathbf{\dot{a}_{pn}} = -2\frac{m}{r^3}\dot{r}(A\mathbf{n}+B\mathbf{v})+\frac{m}{r^2}(\dot{A}\mathbf{n}+\dot{B}\mathbf{v}
+A(\mathbf{v}/r-\mathbf{n}\dot{r}/r)+B\mathbf{a})
\end{equation}
where $\dot{A}$ and $\dot{B}$ are the time derivatives of the coefficients $A$ and $B$, which are the summations of

\begin{equation}
\dot{A}_1 = -2(2+\eta)\frac{m}{r^2}\dot{r}-2(1+3\eta)\mathbf{v}\cdot\mathbf{a}+3\eta\dot{r}\ddot{r},
\end{equation}
\begin{align}
\dot{A}_2 &=\frac{3}{2}(12+29\eta)\frac{m^2}{r^3}\dot{r}-4\eta(3-4\eta)v^2\mathbf{v}\cdot\mathbf{a}\nonumber\\
&-\frac{15}{2}\eta(1-3\eta)\dot{r}^3\ddot{r}+\frac{1}{2}\eta(13-4\eta)\frac{m}{r}\bigg(2\mathbf{v}\cdot\mathbf{a}-\frac{v^2\dot{r}}{r}\bigg)\nonumber\\
&+(2+25\eta+2\eta^2)\frac{m}{r}\bigg(2\dot{r}\ddot{r}-\frac{\dot{r}^3}{r}\bigg) \nonumber \\
&+3\eta(3-4\eta)(\mathbf{v}\cdot\mathbf{a}\dot{r}^2+v^2\dot{r}\ddot{r}),
\end{align}
\begin{align}
\dot{A}_{5/2}&=\frac{8}{5}\eta\frac{m}{r}\bigg(\ddot{r}-\frac{\dot{r}^2}{r}\bigg)\bigg(\frac{17}{3}\frac{m}{r}+3v^2\bigg)\nonumber\\
&+\frac{8}{5}\eta\frac{m}{r}\dot{r}\bigg(-\frac{17}{3}\frac{m}{r^2}\dot{r}+6\mathbf{v}\cdot\mathbf{a}\bigg)
\end{align}
and 
\begin{equation}
\dot{B}_1 =2(2-\eta)\ddot{r},
\end{equation}
\begin{align}
\dot{B}_2 &= -\frac{1}{2}(4+41\eta+8\eta^2)\frac{m}{r}\bigg(\ddot{r}-\frac{\dot{r}^2}{r}\bigg) \nonumber \\
&+\frac{1}{2}\eta(15+4\eta)(2\mathbf{v}\cdot\mathbf{a}\dot{r}+v^2\ddot{r})-\frac{9}{2}\eta(3+2\eta)\dot{r}^2\ddot{r},
\end{align}
\begin{equation}
\dot{B}_{5/2} =\frac{8}{5}\eta\frac{m}{r^2}\dot{r}\bigg(3\frac{m}{r}+v^2\bigg)
-\frac{8}{5}\eta\frac{m}{r}\bigg(-3\frac{m}{r^2}\dot{r}+2\mathbf{v}\cdot\mathbf{a}\bigg)
\end{equation}
where $\ddot{r}$ is the second time derivative of the radial distance given by
\begin{equation}
\ddot{r} = (v^2+\mathbf{r}\cdot\mathbf{a}-\dot{r}^2)/r.
\end{equation}

\label{lastpage}
\end{document}